\documentclass[11pt]{article}

\usepackage{amsmath,mathrsfs,amsbsy,amsthm}
\usepackage{cite}
\usepackage{graphicx}
\usepackage{latexsym,amssymb}
\usepackage{ifpdf,lineno}
\usepackage{color}
\usepackage{geometry}
\usepackage{accents}

\begin{document}
	
	\title{{\bf The Possibility of Tachyons as Soliton Solutions  in $3+1$ Dimensions via  $k$-field Models}}
	
\author{M. Mohammadi \\
	{\small \texttt{physmohammadi@pgu.ac.ir}}}
	\date{{\em{Physics Department, Persian Gulf University, Bushehr 75169, Iran.}}}
	\maketitle

\begin{abstract}

Tachyons or hypothetical faster-than-light (FTL) particles  would fail the principle of causality. Such particles may only be imagined when they have no energy and momentum and, thus, no observable  interaction. 
In this paper,  we show that the classical relativistic field theory with non-standard Lagrangian densities ($k$-field models) does not rule out the existence of FTL particle-like solutions (solitons) with zero energy and momentum in $3+1$ dimensions.  
However, contrary to our expectation, we show that it is possible to have $k$-field models that lead to energetic FTL soliton solutions in $3+1$ dimensions. 

\end{abstract}


\section{Introduction}\label{sec1}

In 1905, Albert Einstein presented his special theory of relativity \cite{Albert}. One of the most important results of this theory was the violation of the causality principle   for faster-than-light (FTL)\footnote{This article uses the terms  “FTL’’ and  “superluminal’’ for the faster-than-light speeds in the vacuum, i.e., $v>c$. } speeds. 
Indeed, the universal speed of light ($c$) was proposed as  an intrinsic limit for particles and in all equations of physical models. For example, any particle with a non-zero rest mass needs infinite energy to reach the speed of light, meaning that particles with non-zero rest masses can only be found at sub-light speeds ($v<c$).  Or, concerning the theories of electromagnetics and gravity, we expect the speed of electromagnetic and gravitational waves not to be greater than that of light. 

Despite the  claims of standard special relativity, the possibility and implication of the hypothetical FTL particles have been an issue of interest among physicists \cite{Bilaniuk,Feinberg,Fox,Camenzind,Asaro,Recami,Liberati,Hill}. 
 Gerald Feinberg \cite{Feinberg} first called these hypothetical particles tachyons and considered them  the quanta of a  relativistic quantum field theory with imaginary rest mass.  Asaro \cite{Asaro} considered complex speeds for  special relativity and  provided meaningful formulae for tachyons. In  \cite{Hill}, two possible new transformations between inertial frames as complementary to the Lorentz
 transformation for regime $c<v<\infty$ were proposed without the need to introduce 
  imaginary masses and complex speeds. 
 All the previous works agree that the speed of light is a limit on both sides,  which means that the speed of subluminal  and (hypothetical) superluminal  particles cannot exceed the speed of light from below and above. 
Despite many qualms in the physics community and the lack of direct experimental evidence, tachyons have been considered in some physical models. For example,  it has been claimed that some neutrinos may be  tachyons \cite{Ciborowski,Ehrlich,Schwartz2}. In cosmology,  tachyons  were proposed as a way to  model  cosmic inflation and dark energy \cite{Feinstein,Hussain}.

%In cosmology  tachyons  are used as a way of explaining cosmic inflation and dark energy \cite{Feinstein}.

In the framework of standard special relativity for subluminal particles ($v<c$), the usual expressions  for mass, energy, and momentum   are  
\begin{equation} \label{st}
	m=\dfrac{m_{o}}{\sqrt{1-v^2/c^2}},\quad E=mc^2, \quad |\textbf{P}|=mv,
\end{equation}
where $m_{o}$ denotes the rest-mass, and $E^2=|\textbf{P}|^2c^2+(m_{o}c^2)^2$ is known  as the energy-momentum relation. 
Thus as $v\longrightarrow c$ from below, $|\textbf{P}|$  and $E$ become infinite
and would become imaginary for $v>c$. For photons with $m_{o}\longrightarrow 0$, the speed $v$ can tend to the speed of light so that the total energy tends to a finite non-zero value $E=|\textbf{P}|c$. It should be noted that if the rest mass is absolutely zero, i.e., $m_{o}=0$, the energy is zero for all speeds, even for $v\longrightarrow c$.  
A standard way to apply usual kinematic relations (\ref{st}) to hypothetical superluminal particles ($v>c$)  is to assume that they are entities with complex rest masses, i.e., $m_{o}=iM$, where $M$ is real \cite{Feinberg}. Hence, the energy and momentum will remain real quantities, satisfying:  
\begin{equation} \label{stf}
	 E=\dfrac{M c^2}{\sqrt{v^2/c^2-1}}, \quad |\textbf{P}|=\dfrac{M v}{\sqrt{v^2/c^2-1}}=\dfrac{E}{c^2}v.
\end{equation}
Also, the energy-momentum relation would be  $E^2=|\textbf{P}|^2c^2-(M c^2)^2$.
Interestingly, energy and momentum are monotonic
decreasing functions of the speed so that for $v=\infty$, we have $E=0$ and $|\textbf{P}|=M c$. That is, for infinite speed, tachyons carry momentum but no energy. 
It should be noted that  Einstein's usual velocity  transformation relation remains valid   for the FTL speeds as well. Hence, infinite speed is not a  Lorentz invariant \cite{Feinberg,Hill}.

The present article  examines the possibility of FTL soliton solutions in $3+1$ dimensions, which  has been previously  done in $1+1$ dimensions \cite{Mohammadi1}. In the theory of relativistic classical  fields, solitons are special solutions of field equations that have a localized energy density form and are stable against any deformation \footnote{Typically, a soliton is a stable solitary wave solution without distortion after collisions. However, this paper only considers  the stability condition for defining  solitons.} \cite{Lamb,Rajaraman,Manton}.  Similar to real particles, they satisfy the well-known relativistic energy-momentum relations of  special relativity (\ref{st}), and their dimensions contract in the direction of motion according to the Lorentz contraction law. 
In general, solutions with localized energy density (solitary wave solutions) can be classified into topological and non-topological categories. Topological solitary wave solutions are inherently  stable and are solitons, but non-topological ones are not necessarily stable. The kink-bearing models in  $1+1$ dimensions \cite{Campbell1986,Anninos1991,Goodman2005,Dorey2011,Bazeia2018,Gani2019,Bazeia2019,Mohammadi2022,Campos2023,Evslin2023,Pereira2023},  Skyrme’s model of baryons \cite{Manton,SKrme,SKrme2,SKrme3,SKrme4},
and ’t Hooft-Polyakov’s model for monopole  \cite{Rajaraman,Manton,toft,pol,MKP,TOF,CTOPO}  are instances of field models with topological solitons. Despite the variety of topological solitons, the only well-known non-topological relativistic solitons are Q-balls \cite{Rosen,Marques,Friedberg,Bazeia2016,Panin,R2,R3} that have been proposed as dark matter candidates \cite{Kusenko,Shoemaker,Cotner,Gorbunov}.

Regarding the principle of causality, it is clear that the assumption of an FTL particle that carries energy and momentum when interacting with nature violates the principle of causality. Nevertheless, the assumption of FTL particles with zero energy and momentum, which naturally has no  observable interactions, does not contradict the principle of causality. Such hypothetical  particles can be produced in infinite numbers without spending energy. From an energy point of view, the only possible  effect of the existence of such  particles in a system is causing  it to change from one state to another with the same energy; that is, the changes between degenerate energy levels (provided other conservation laws are not violated).

In the relativistic classical field theory framework, a special type of non-standard Lagrangian (NSL) densities should be used to obtain a zero-energy particle-like solitary wave solution \cite{Mohammadi1,R3,MMQ1,MMQ4}. In general, for several relativistic scalar fields $\Phi_{r} (r=1\cdots N$), an NSL density is defined as a non-linear functional of the kinetic scalar terms ${\cal S}_{ij}=\partial_{\mu}\Phi_{i}\partial^{\mu}\Phi_{j}$. Equivalently, the  term $k$-field  has been coined for such a system of relativistic fields whose Lagrangian density  is not linear in ${\cal S}_{ij}$.  
The special  solitary wave solutions of $k$-fields are known as defect structures \cite{R2,Baz1,Adam,Bab}. Many papers, such as  \cite{Baz1,Adam,Bab,Vash,Mahzon}, deal  with such systems with defect structures (e.g., domain walls, vortices, and monopoles). In cosmology, such models are usually called $k$-essence \cite{Armendarizn,Chiba2,Rendall,Armend3} and have gained particular popularity. They are
used to describe dark energy and dark matter \cite{Armend3,cos1,cos2,cos3,cos4} or are  proposed in the context of inflation leading to $k$-inflation \cite{Armend,Chiba,Armend2,cosmology1}. 
Although the primary  goal of this paper is to introduce a zero-energy FTL soliton in $3+1$ dimensions, we show that  the existence of  energetic  FTL  solitons is also possible. 
Overall, the classical relativistic field theory in $3+1$ dimensions is not inherently incompatible with the existence of particle-like FTL soliton solutions.

The organization of this paper is as follows: in the next
section, the $k$-field systems that lead to  zero-energy solutions
will be generally introduced. In section \ref{sec3}, through two examples, we describe the general rule for achieving an  FTL solitary wave solution. Section \ref{sec4} presents  a $k$-field model that leads to a single zero-energy FTL soliton solution. A modified version of such a $k$-field model will be introduced in section \ref{sec5} and will yield an energetic FTL soliton solution. Section \ref{sec6} examines the stability of the FTL soliton solution obtained in the previous section. The last section is devoted to a summary and conclusion.

%, the general form of the classical Lagrangian densities that yields to zero-energy solitary wave solutions will be introduced .
%, for several relativistic scalar fields $\Phi_{r} (r=1\cdots N$), the general form of the classical Lagrangian densities that yields to zero-energy solitary wave solutions will be introduced . 
%Such Lagrangian densities  are obtained as a non-linear functional of the kinetic scalar terms ${\cal S}_{ij}=\partial_{\mu}\Phi_{i}\partial^{\mu}\Phi_{j}$.   In general, the term $k$-fields has been coined for  systems of relativistic fields whose Lagrangian densities are not linear in ${\cal S}_{ij}$. 

% regarding the impossibility of FTL speed and causality principle

\section{Zero-Energy  Solutions}\label{sec2}

The Lagrangian densities of the classical relativistic field theories are introduced as functionals of  some fields and their first derivatives:
\begin{equation} \label{lag}
	{\cal L}= {\cal L}(\Phi_{r},\Phi_{r,\mu}),
\end{equation}
where $\Phi_{r,\mu}=\frac{\partial \Phi_{r}}{\partial x^{\mu}}$ and $x^{\mu}\equiv (ct,x,y,z)$. The index $r$  is used for labeling  different independent scalar fields and  can be used to label the component of each vector field (for example, the electromagnetic  potential field $A^{\mu}$). The allowed combinations of the relativistic fields $\Phi_{r}$ and their first derivatives $\Phi_{r,\mu}$ in the Lagrangian density (\ref{lag}) should be  such that it remains a scalar functional.  According to the principle of least action, the Euler-Lagrange  dynamical  equations  would be:
\begin{equation} \label{lfg}
	\frac{\partial{\cal L}}{\partial\Phi_{r}}-\frac{\partial}{\partial x^{\mu}}\left(\frac{\partial {\cal L}}{\partial \Phi_{r,\mu}}\right)=0.
\end{equation}
The Lagrangian density (\ref{lag}) is invariant under the infinitesimal space-time translations. For infinitesimal translations, using Noether's theorem, four separate continuity equations $\partial_{\mu}T^{\mu\nu}=0$ and then four  conserved quantities $cP^{\mu}=\int T^{o\mu}d^{3}\textbf{x}$ would  be obtained, where
\begin{equation} \label{egb}
	T^{\mu\nu}=\sum_{r=1}^{N}\frac{\partial{\cal L}}{\partial\Phi_{r,\mu}}\frac{\partial\Phi_{r}}{\partial x_{\nu}}-{\cal L}g^{\mu\nu}
\end{equation}
is called the energy-momentum tensor, and $g^{\mu\nu}$  is   the  Minkowski metric. The energy and momentum components   of the  system (\ref{lag}) are introduced as $cP^{o}$  and $P^{j}$ ($j=1,2,3$), respectively.  Hence,    $T^{00}$ is called the  energy density, and $p^{j}=(1/c)T^{0j}$ ($j=1,2,3$)  are called the components of the momentum density:
\begin{equation} \label{e5b}
	T^{00}=\epsilon=\sum_{r=1}^{N}\Pi_{r}\dot{\Phi}_{r}-{\cal L},
\end{equation}
\begin{equation} \label{e5a}
	T^{0j}=-\sum_{r=1}^{N}c~\Pi_{r}\frac{\partial\Phi_{r}}{\partial x^{j}}.
\end{equation}
where $\Pi_{r}=\frac{\partial{\cal L}}{\partial\dot{\Phi}_{r}}$ are the conjugate fields,  and $\dot{\Phi}_{r}=\frac{\partial\Phi_{r}}{\partial t}$.

We can introduce a zero-energy solution as a non-trivial solution of Eq.~(\ref{lfg}) whose energy density is zero everywhere. If we assume  there are $N$ independent fields ${\Phi}_{r}$ (i.e., $r=1,2,\cdots,N$), condition $\epsilon(x,t)=0$ can be assumed  as a new partial differential equation (PDE)  along with $N$ coupled PDEs (\ref{lfg}). Typically,   it  is unlikely that $N+1$ PDEs for $N$ unknown  fields $\Phi_{r}$ lead to a common (joint) solution. However, a zero-energy solution can be obtained if the Lagrangian density ${\cal L}$ and all its derivatives (i.e.  $ \frac{\partial{\cal L}}{\partial\Phi_{r}}$, and $\frac{\partial {\cal L}}{\partial \Phi_{r,\mu}}$) vanish for that solution simultaneously, and those $N+1$ PDEs  will  be satisfied  automatically. Note that such conditions lead to zero momentum as well.

From here onward, we restrict ourselves to systems whose Lagrangian densities are functionals of the scalar fields; that is, they are not functionals of the vector fields.
Therefore, for a set of relativistic  scalar  fields $\Phi_{i}$ ($i=1,\cdots, N$), the  Lagrangian densities  must be functionals   of the fields and the kinetic scalars ${\cal S}_{ij}={\cal S}_{ji}=\partial_{\mu}\Phi_{i}\partial^{\mu}\Phi_{j}$:
\begin{equation} \label{slag}
	{\cal L}= {\cal L}(\Phi_{k},{\cal S}_{ij}),\quad j\leq i,
\end{equation}
where $i,k=1,\cdots, N$. The non-standard Lagrangian densities, which are not linear in terms of the kinetic scalars ${\cal S}_{ij}=\partial_{\mu}\Phi_{i}\partial^{\mu}\Phi_{j}$, are called $k$-fields.  
The dynamical  equations   (\ref{lfg}) for such systems  (\ref{slag}) are simplified  to
\begin{equation} \label{lfg2}
	\frac{\partial{\cal L}}{\partial\Phi_{r}}-\sum_{j=1}^{N}(1+\delta_{rj})\left[\frac{\partial}{\partial x^{\mu}}\left(\frac{\partial {\cal L}}{\partial {\cal S}_{rj}}\right)\partial^{\mu}\Phi_{j}+\frac{\partial{\cal L}}{\partial {\cal S}_{rj}}\partial_{\mu}\partial^{\mu}\Phi_{j}\right]=0.
\end{equation}
In the same way, the energy density function (\ref{e5b}) is reduced to
\begin{equation} \label{e5b2}
	\epsilon=\sum_{r=1}^{N}\sum_{i=1}^{N}\frac{1}{c^2}\frac{\partial{\cal L}}{\partial{\cal S}_{ir}}\dot{\Phi}_{i}\dot{\Phi}_{r}(\delta_{ir}+1)-{\cal L}.
\end{equation}
Therefore,  to have a zero-energy solution,  ${\cal L}$ and  the following derivatives  $\frac{\partial{\cal L}}{\partial\Phi_{i}}$, as well as $\frac{\partial{\cal L}}{\partial{\cal S}_{ij}}$  should  become zero simultaneously.
To better understand how such a situation is possible, let us use a single scalar field $\Phi$  and  introduce an NSL density as ${\cal L}=\mathbb{S}^3$ in $1+1$ dimensions, where $\mathbb{S}={\cal S}+4\varphi^2\ln |\varphi|$ and ${\cal S}=\partial_{\mu}\varphi\partial^{\mu}\varphi$. It is easy to find $\varphi=\pm\exp(x^2)$  as a static solution for  condition  $\mathbb{S}=0$. Hence, conditions  
${\cal L}=\mathbb{S}^3=0$, $\frac{\partial{\cal L}}{\partial\Phi}=3\mathbb{S}^2\frac{\partial\mathbb{S}}{\partial\Phi}=0$, and $\frac{\partial{\cal L}}{\partial{\cal S}}=3\mathbb{S}^2=0$ would be satisfied automatically for such a static function. In other words, the static function $\varphi=\pm\exp(x^2)$ is a zero-energy solution for the NSL density  ${\cal L}=\mathbb{S}^3$. 
In general, for a system of  scalar fields  $\Phi_{i}$ ($i=1,\cdots, N$), any arbitrary number of the independent scalar functionals    $\mathbb{S}_{j}$ ($j=1,\cdots,m$),  which are all  zero simultaneously ($\mathbb{S}_{j}=0$) for a special solution,  can be used to introduce  an NSL  density (a $k$-field model) with  a zero-energy solution as follows:
\begin{eqnarray} \label{sf0}
	{\cal L}=\sum_{n_{1}=0}^{\infty}\sum_{n_{2}=0}^{\infty}\cdots\sum_{n_{m}=0}^{\infty} a({n_{1},\cdots,n_{m}})\mathbb{S}_{1}^{n_{1}}\mathbb{S}_{2}^{n_{2}}\cdots\mathbb{S}_{m}^{n_{m}},
\end{eqnarray}
provided $n_{1}+n_{2}+\cdots+n_{m} \geqslant 3$.    Note that,  scalars  $\mathbb{S}_{j}$ ($j=1,\cdots,m$) and  coefficients $a({n_{1},\cdots,n_{m}})$  can be arbitrary    well-defined functionals  of the fields and the kinetic scalars ${\cal S}_{ij}$.
Accordingly, constructing a $k$-field model with a zero-energy solution seems to be simple.
However, expecting a system to  have a single zero-energy solution involves a lot of complications. For example, the earlier system ${\cal L}=\mathbb{S}^3$ does not yield a single zero-energy solution in $3+1$ dimensions. 
In fact, it has a variety of zero-energy solutions  $\exp({-(r-\xi)^2})$, where $\xi$ can be any arbitrary number.  Furthermore, in line with the objectives of this article,
all terms in the energy  density of the constructed  $k$-field model  should be positive definite.
To be more precise, we want to introduce a $k$-field model with a single zero-energy solution that is energetically stable, 
and a solution is called energetically stable if any variations above its background lead to an increase in the total energy.

\section{How to obtain the relativistic FTL solutions} \label{sec3}

In general, for a  relativistic scalar field model (\ref{slag}), if  $\Phi_{ro}(x^o,x^1,x^2,x^3)$ is a  non-moving  solution, the moving version of that which moves at velocity  $\textbf{v}$ can be obtained easily by a relativistic boost  $\Phi_{rv}=$ $\Phi_{ro}$ $(\Lambda^{0}_{~\nu}$$ x^{\nu},\Lambda^{1}_{~\nu}$$ x^{\nu},\Lambda^{2}_{~\nu} x^{\nu},\Lambda^{3}_{~\nu} x^{\nu})$, where  $\Lambda^{\mu}_{~\nu}(\textbf{v})$ are   the matrix  elements  of the Lorentz transformation restricted by the conditions $\Lambda^{\mu}_{~\nu}\Lambda^{\alpha}_{~\beta}g_{\mu\alpha}=g_{\nu\beta}$. There would be the same standard relativistic energy-momentum relations (\ref{st}) between the moving and non-moving versions of  any localized solution in which the components of the energy-momentum tensor $T^{\mu\nu}$ symptomatically approach zero at infinity. In other words,  if $E_{v}=cP^{o}$ and $\textbf{P}_{v}=(P^{1},P^{1},P^{3})$ are  the total energy and momentum of a  localized moving solution of a relativistic  field system (\ref{lag}), we have: $E_{v}=\gamma E_{o}$ and $\textbf{P}_{v}=\gamma E_{o} \textbf{v}/c^2$, where $v=|\textbf{v}|$, $\gamma=1/\sqrt{1-(v/c)^2}$, and  $E_{o}$ is the rest energy. Note that  the total energy of a zero-energy solution is always zero and is independent of velocity.

For example, for a complex scalar field $\phi$ in $3+1$ dimensions,  the following standard Lagrangian density can be considered \cite{Marques}:
 \begin{equation} \label{lagexam}
	{\cal L}= \partial_\mu \phi^*\partial^\mu \phi -|\phi |^2(2-\ln|\phi |^2).
\end{equation}
The related field equation would be obtained as
\begin{equation} \label{eq}
	\Box \phi =\frac{1}{c^2}\frac{\partial^2\phi}{\partial
		t^2}- \nabla^{2} \phi=\phi(-1 + \ln(|\phi|^2)).
\end{equation}
This equation leads to  infinite non-moving spherically
symmetric Q-ball solutions, which characterized by different rest
frequencies $\omega_{o}$:
\begin{equation} \label{f1}
	\phi_{o}(x,y,z,t)=A(\omega_{o})\exp{\left(-\frac{x^2+y^2+z^2}{2}\right)}\exp(i\omega_{o}t),
\end{equation}
where  $A(\omega_{o})=\exp(2-\omega_{o}^2/2c^2)$, and $0\leqslant |\omega_{o}|\leqslant \infty$.  Hence, a moving  Q-ball
solution, which, e.g.,  moves in the $x$-direction
with a constant velocity $\textbf{v}=v\hat{i}$, can be obtained by a relativistic boost ($x\rightarrow \gamma (x-vt)$ and $t\rightarrow \gamma (t-vx/c^2)$):
\begin{equation} \label{es}
	\phi_{v}(x,y,z,t)=A(\omega_{o})\exp{\left(-\frac{\gamma^{2}(x-vt)^2+y^2+z^2}{2}\right)}\exp(i\omega_{o}\gamma (t-vx/c^2)).
\end{equation}  
From a mathematical point of view, the assumption of FTL velocities ($v>c$), e.g., in the $x$-direction, leads to a solution as well:
\begin{equation} \label{f4}
	\phi_{v}(x,y,z,t)=A(\omega_{o})\exp{\left(\frac{\Gamma^{2}(x-vt)^2-y^2-z^2}{2}\right)}\exp(\omega_{o}\Gamma (t-vx/c^2)),
\end{equation} 
where $\Gamma=i\gamma=1/\sqrt{(v/c)^2-1}$. It is clear that the above solution has no physical importance because it is not localized and has infinite energy. Nevertheless, this example illustrates our main idea of how to achieve an FTL solution. Suppose it is possible to find a non-moving solution of a relativistic field system in such a way that its FTL version is localized and has finite energy. In that case, it can be claimed that we have an FTL solitary wave solution.

Another simple example that well provides the possibility of the existence of an FTL solitary wave solution (albeit unstable) in $1+1$ dimensions is introduced by the following Lagrangian density:
 \begin{equation} \label{b}
	{\cal L}=\partial_\mu \varphi\partial^\mu \varphi -U(\varphi),
\end{equation} 
where $\varphi$  is a real scalar field and $U(\varphi)=4(\varphi^4-\varphi^3)$. The corresponding dynamical equation would be 
\begin{equation} \label{jn}
	\Box \varphi =\frac{1}{c^2}\frac{\partial^2\varphi}{\partial
		t^2}- \frac{\partial^2\varphi}{\partial
		x^2}=-8\varphi^3+6\varphi^2,
\end{equation} 
for which there is a non-moving non-localized solution   as follows:
\begin{equation} \label{ng}
\varphi_{o}=\dfrac{1}{1-x^2}.
\end{equation} 
Using an   FTL relativistic boost ($v>c$): $-x^2\rightarrow \Gamma^2 (x-vt)^2$,  a  localized (solitary wave) solution  is obtained:
\begin{equation} \label{bh}
	\varphi_{v}=\dfrac{1}{1+\Gamma^2 (x-vt)^2}.
\end{equation} 
For the Lagrangian density (\ref{b}), the corresponding energy and momentum density  would be:
\begin{equation} \label{bn}
	\varepsilon(x,t)=\frac{1}{c^2}\dot{\varphi}^2+\varphi'^2+U(\varphi),\quad \textrm{and} \quad p(x,t)=-\frac{2}{c^2}\dot{\varphi}\varphi'.
\end{equation}
Thus, the total energy and momentum of the FTL solitary wave solution (\ref{bh}) are obtained as follows:
\begin{equation} \label{vbm}
	E_{v}=\int_{-\infty}^{\infty}[\frac{1}{c^2}\dot{\varphi_{v}}^2+\varphi_{v}'^2+
	U(\varphi_{v})]dx=\frac{\pi}{2}\Gamma=\dfrac{\frac{\pi}{2}}{\sqrt{\frac{v^2}{c^2}-1}}=iM c^2 \gamma,
\end{equation}
and
\begin{equation} \label{ld2}
	P=\int_{-\infty}^{\infty}[-\frac{2}{c^2}\dot{\varphi_{v}}\varphi_{v}']dx=
	(\frac{E_{v}}{c^2}) v,
\end{equation}
where $M=\pi/(2c^2)$. These equations confirm the same  standard  kinematic energy-momentum relations (\ref{stf}) for hypothetical FTL particles.

%Thus, in the regime of FTL speeds, the higher the speed, the lower the energy and momentum. It should be noted that the speed of light is still a limiting speed, meaning that the energy of the FTL solution (\ref{bn}) tends to infinity
% as its speed tends to the speed of light. Also, according to equation (\ref{vbm}), the FTL solitary wave solution (\ref{bh}) can be considered as a particle with imaginary rest mass.  

Therefore, the general rule for obtaining an FTL solution is that we should first find a non-moving solution and then check in which direction the boost of the non-moving solution for FTL velocities turns it into a solitary wave solution. It should be noted that in the right  FTL boost process, the real fields remain real.

\section{A $k$-field model with a zero-energy FTL  soliton solution in $3+1$ dimensions} \label{sec4}

\begin{figure}[htp]
	
	\centering

	\begin{tabular}{cc}

		\includegraphics[width=47mm]{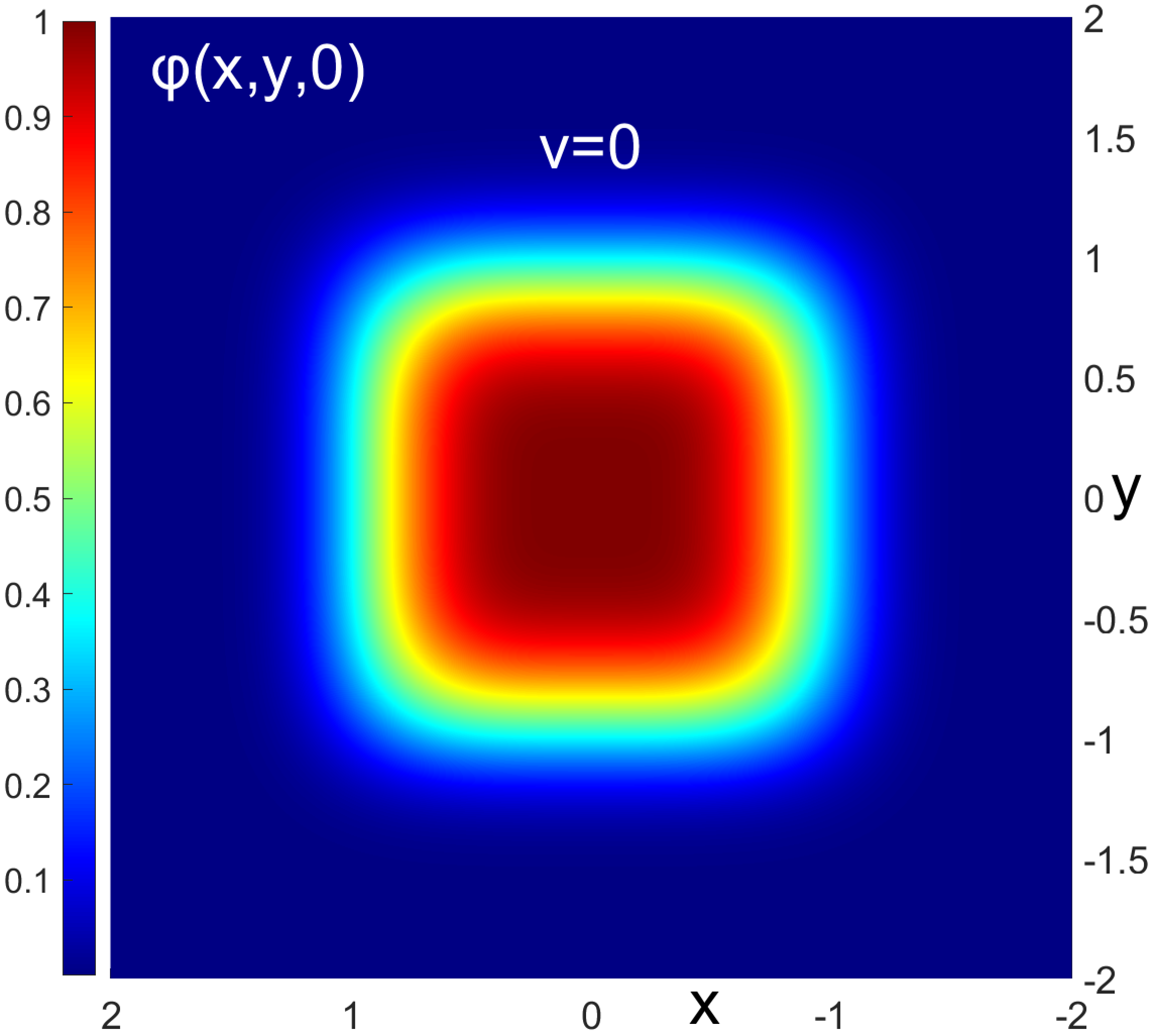}
		
		\includegraphics[width=47mm]{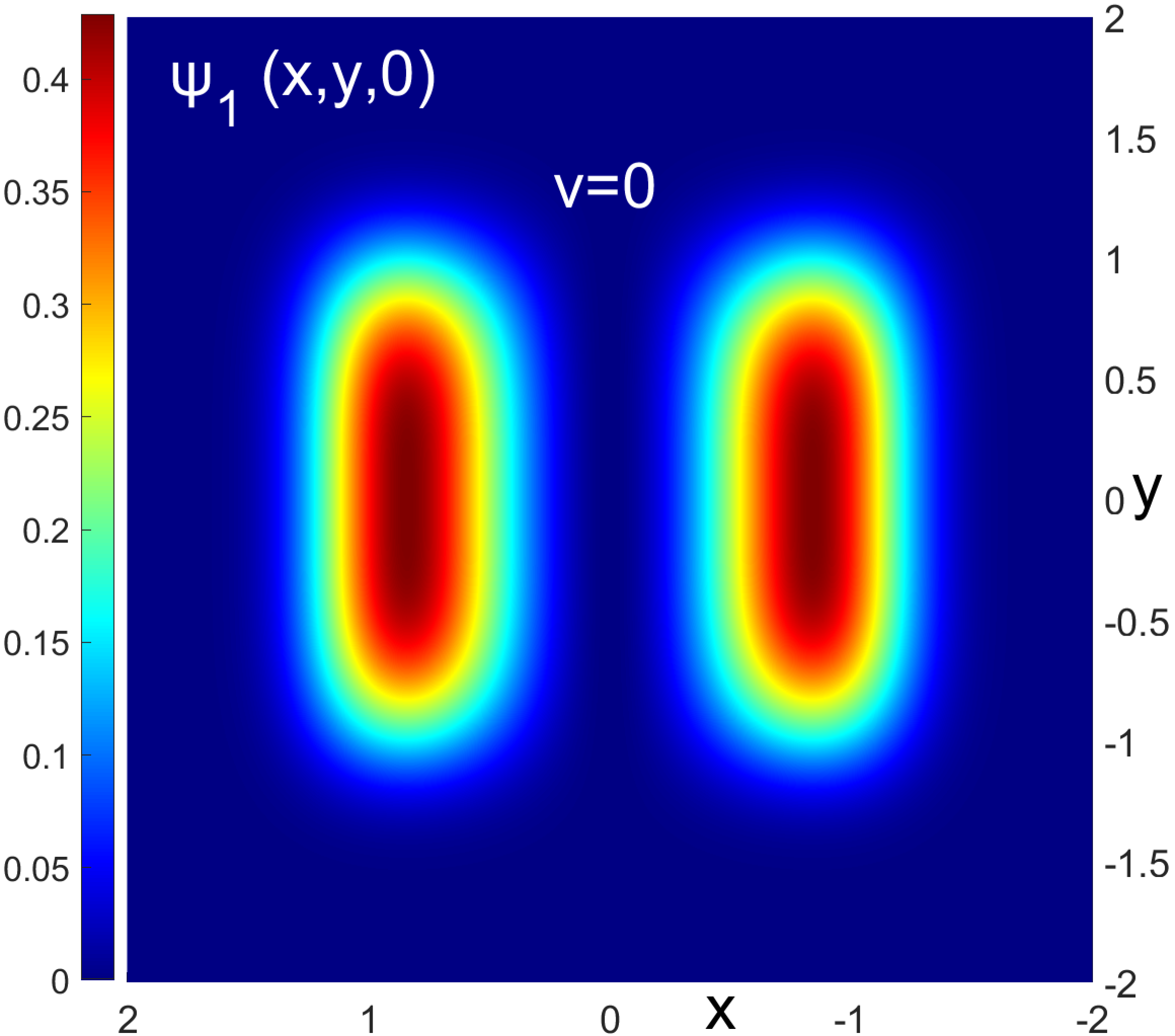}
		
		\includegraphics[width=47mm]{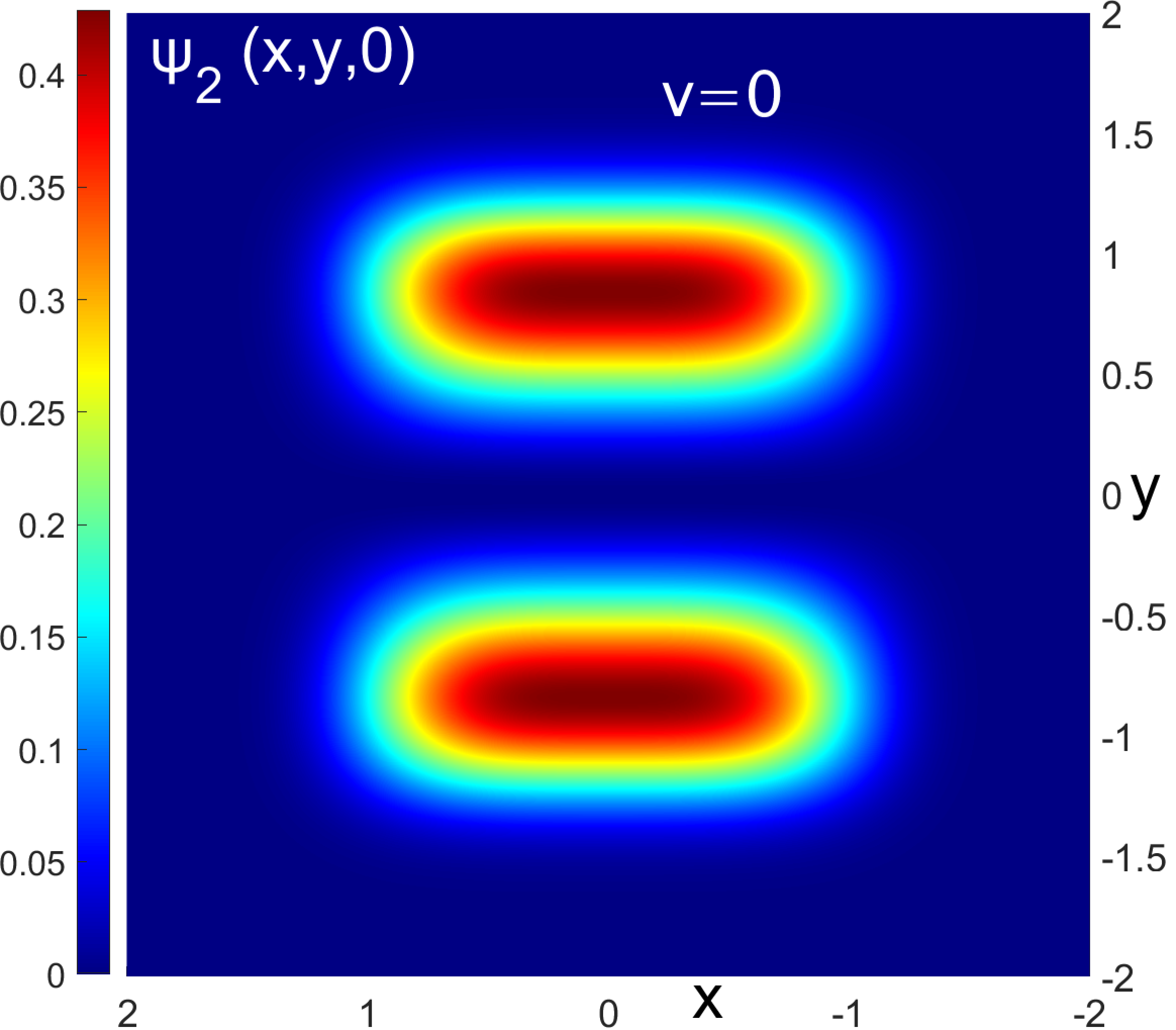}\\
		
		\includegraphics[width=47mm]{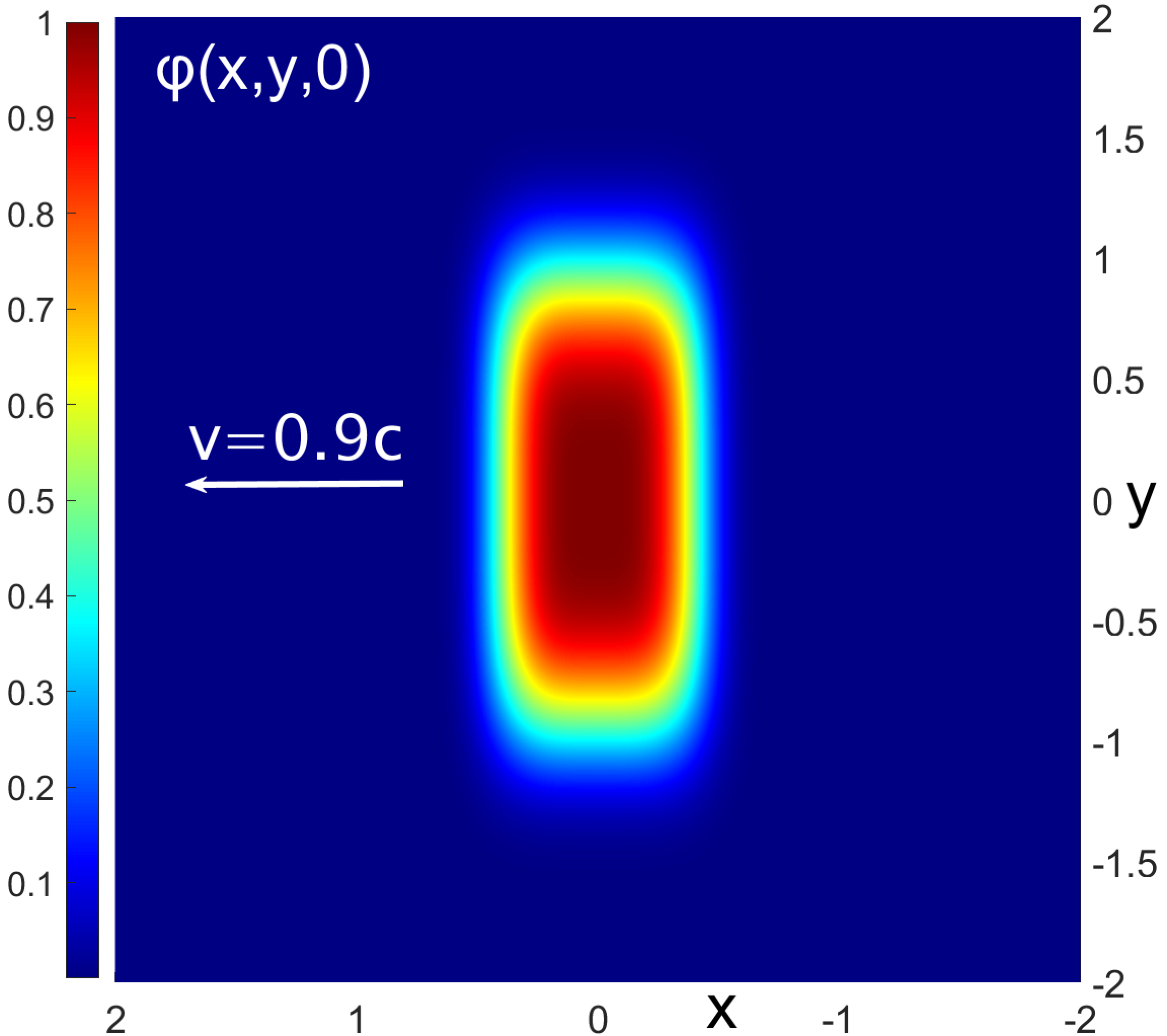}
		
		\includegraphics[width=47mm]{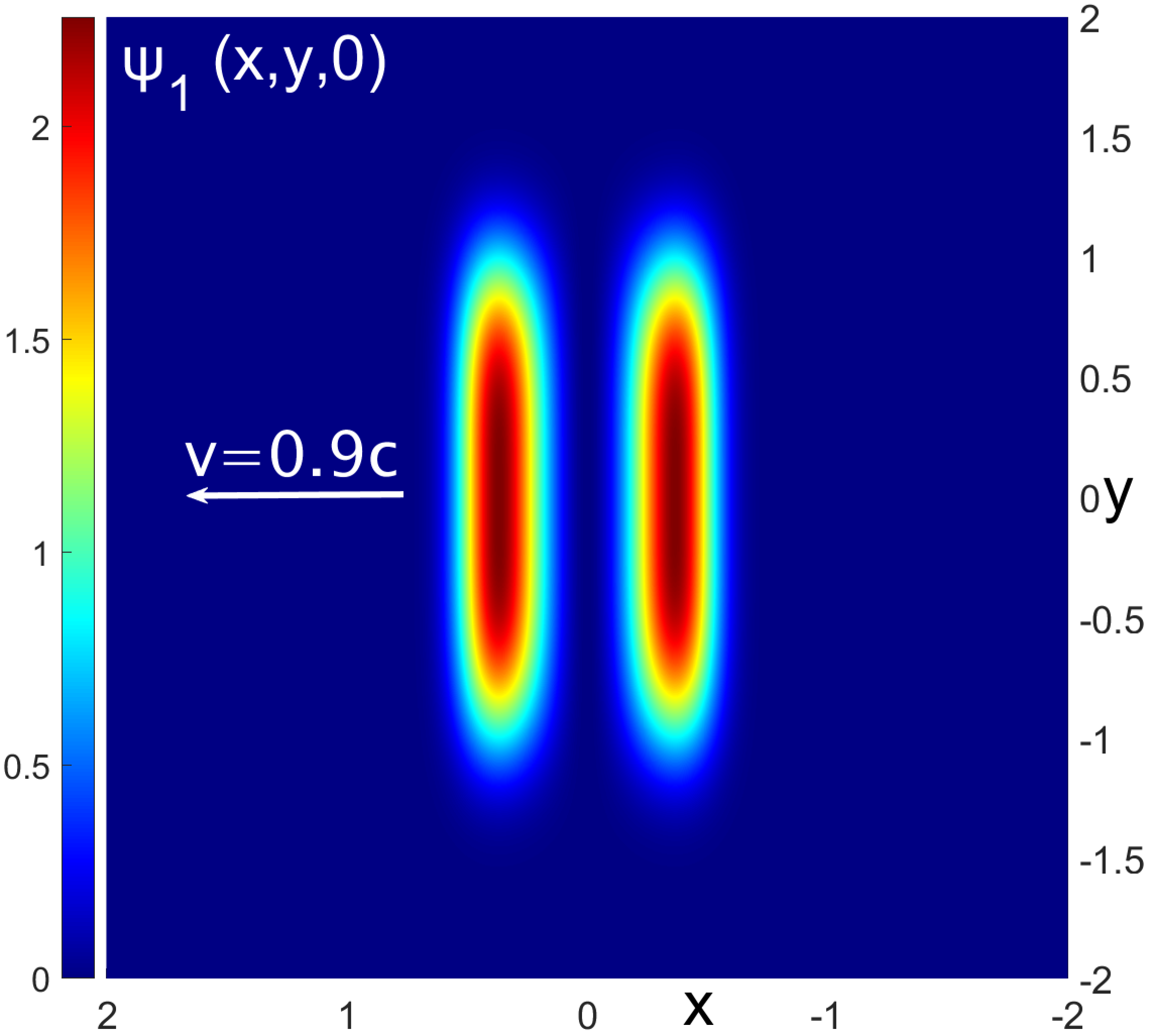}
		
		\includegraphics[width=47mm]{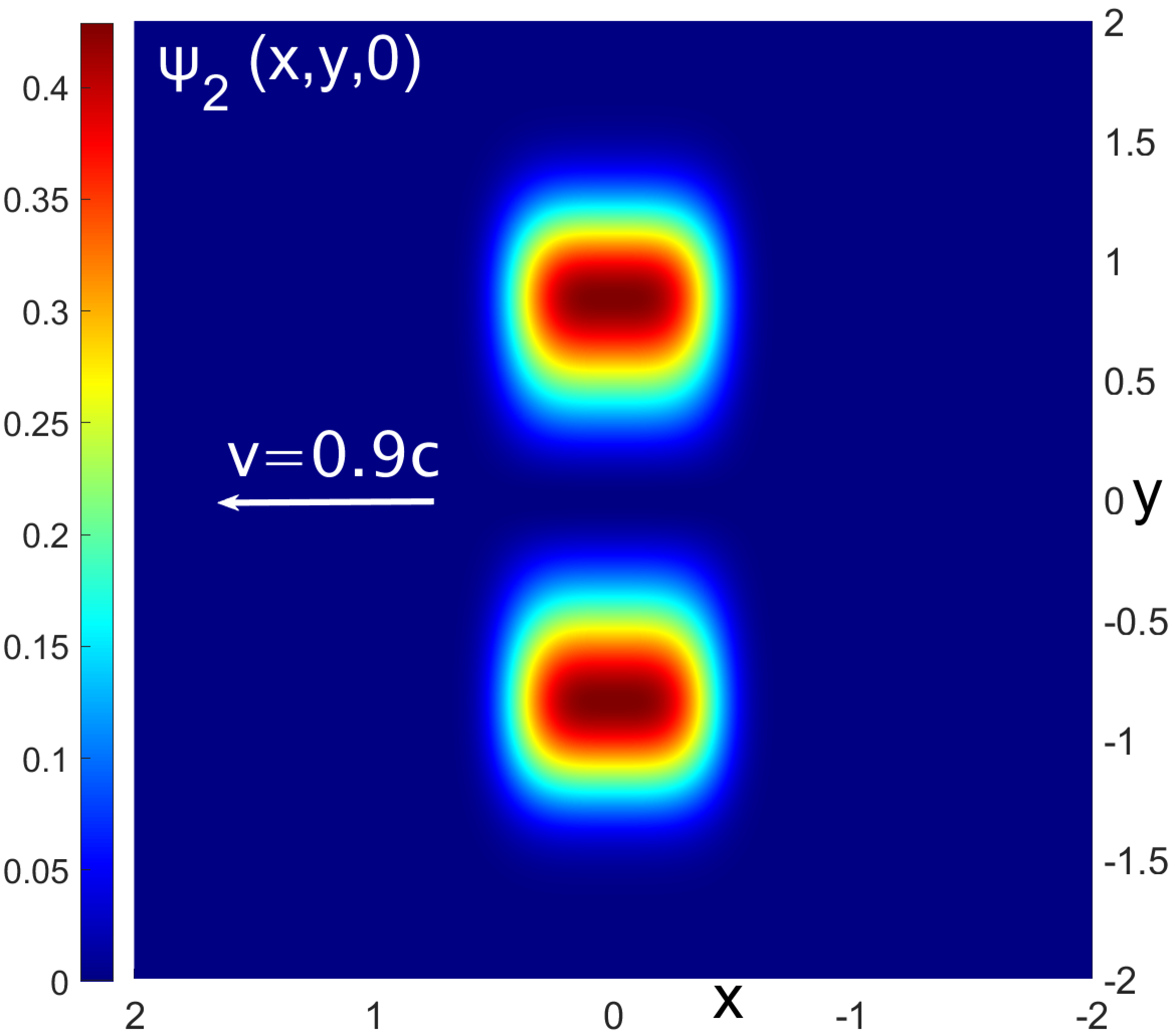}\\
		\includegraphics[width=47mm]{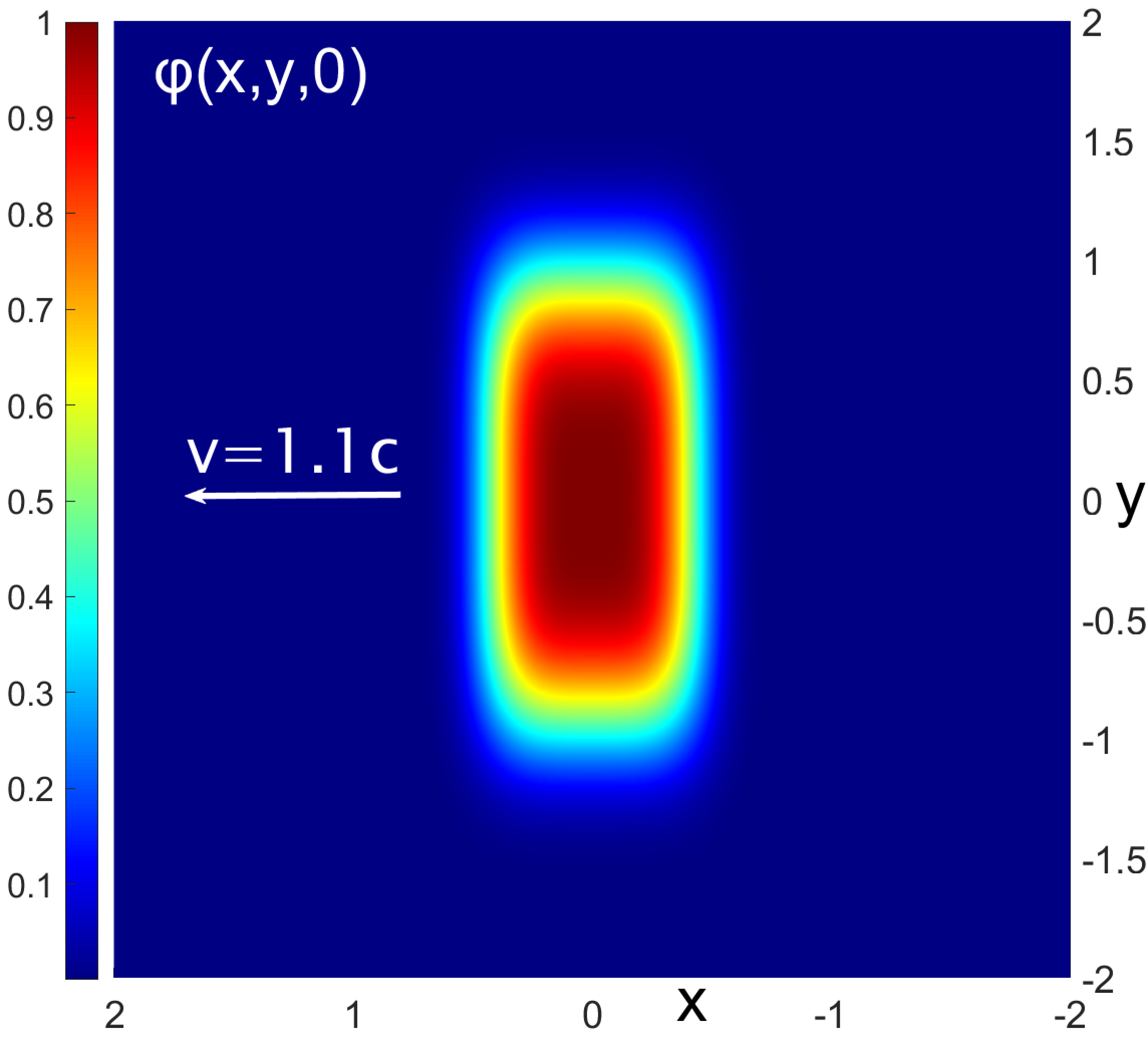}
		
		\includegraphics[width=47mm]{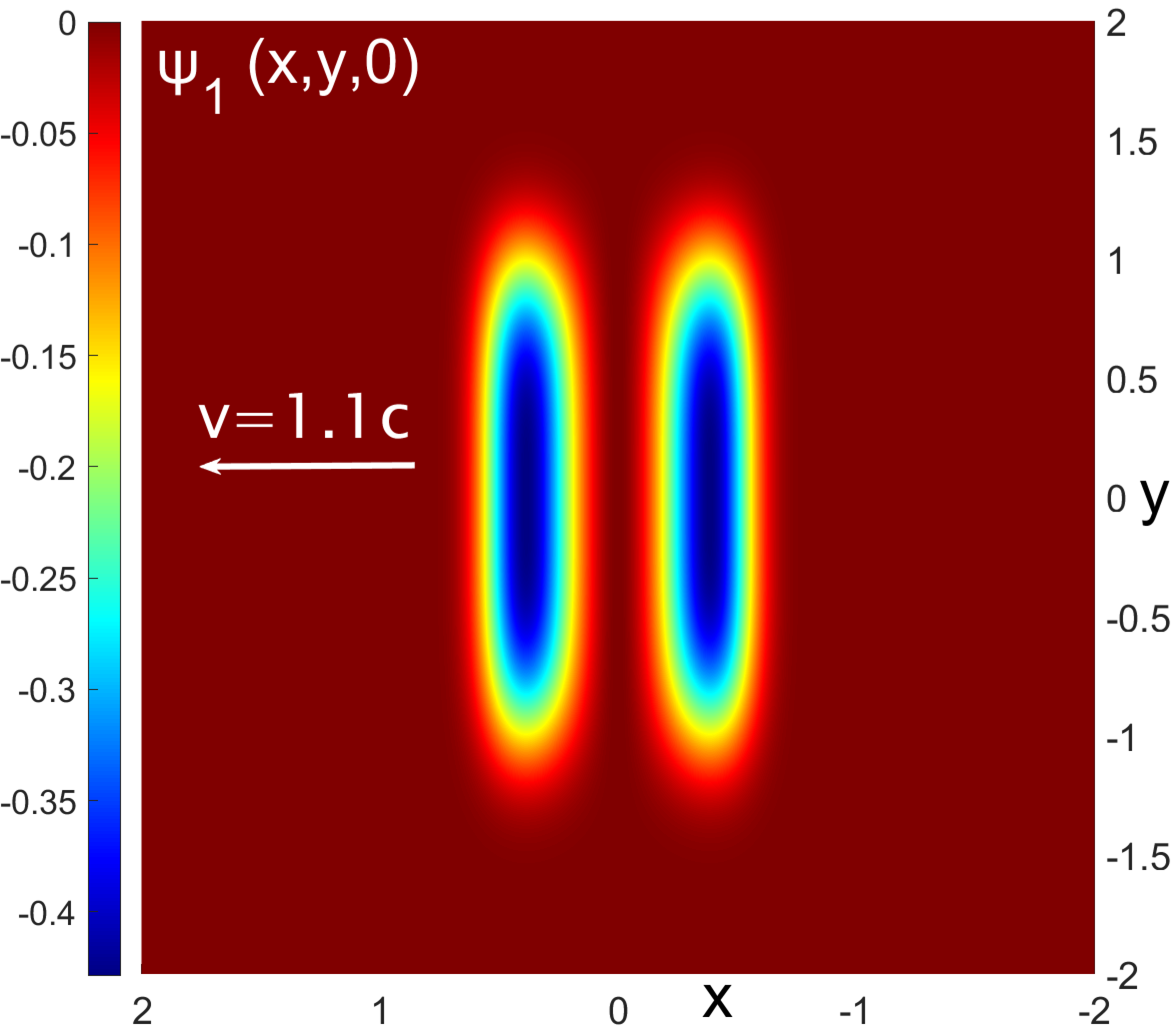}
		
		\includegraphics[width=47mm]{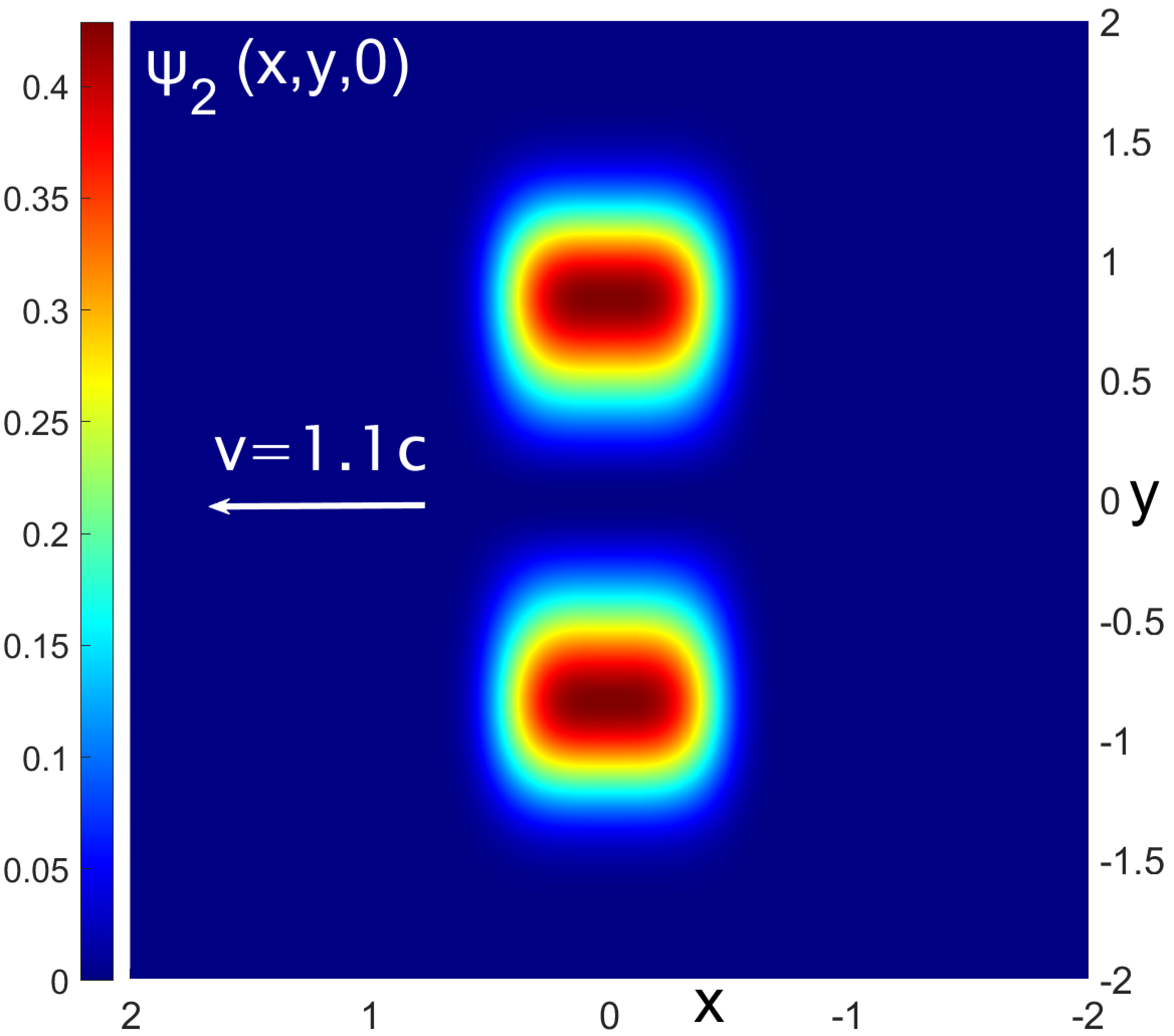}\\
		\includegraphics[width=47mm]{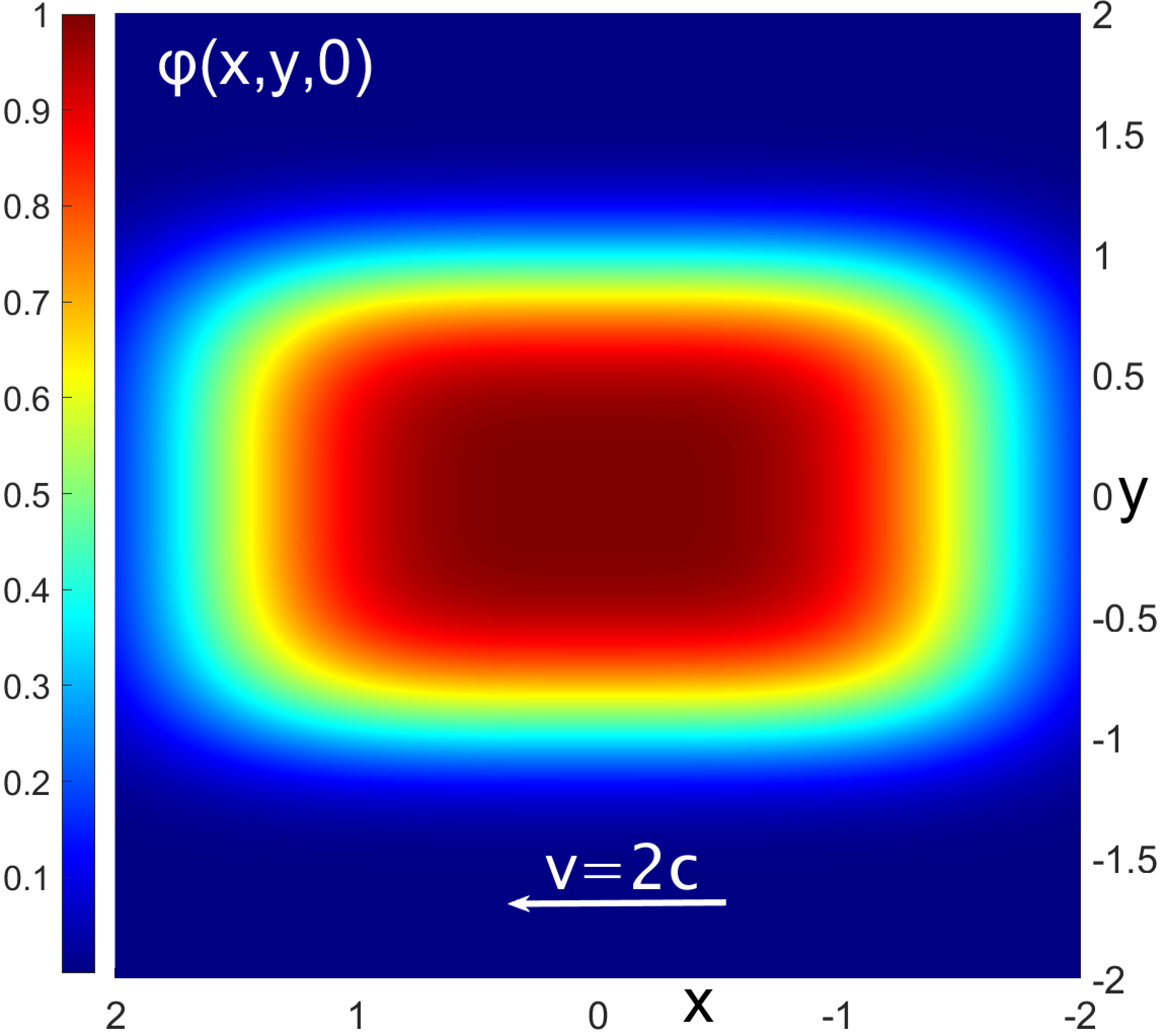}
		
		\includegraphics[width=47mm]{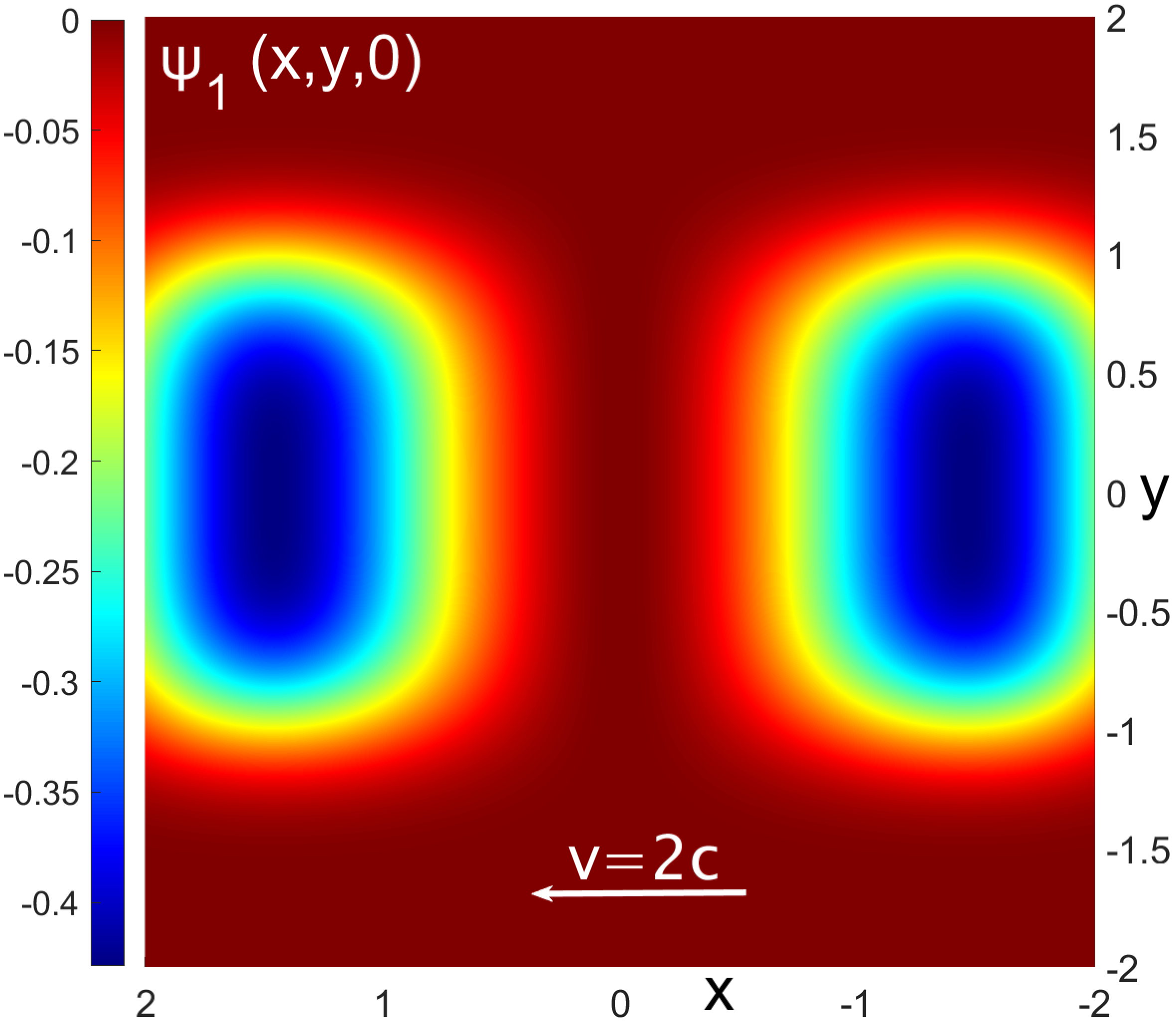}
		
		\includegraphics[width=47mm]{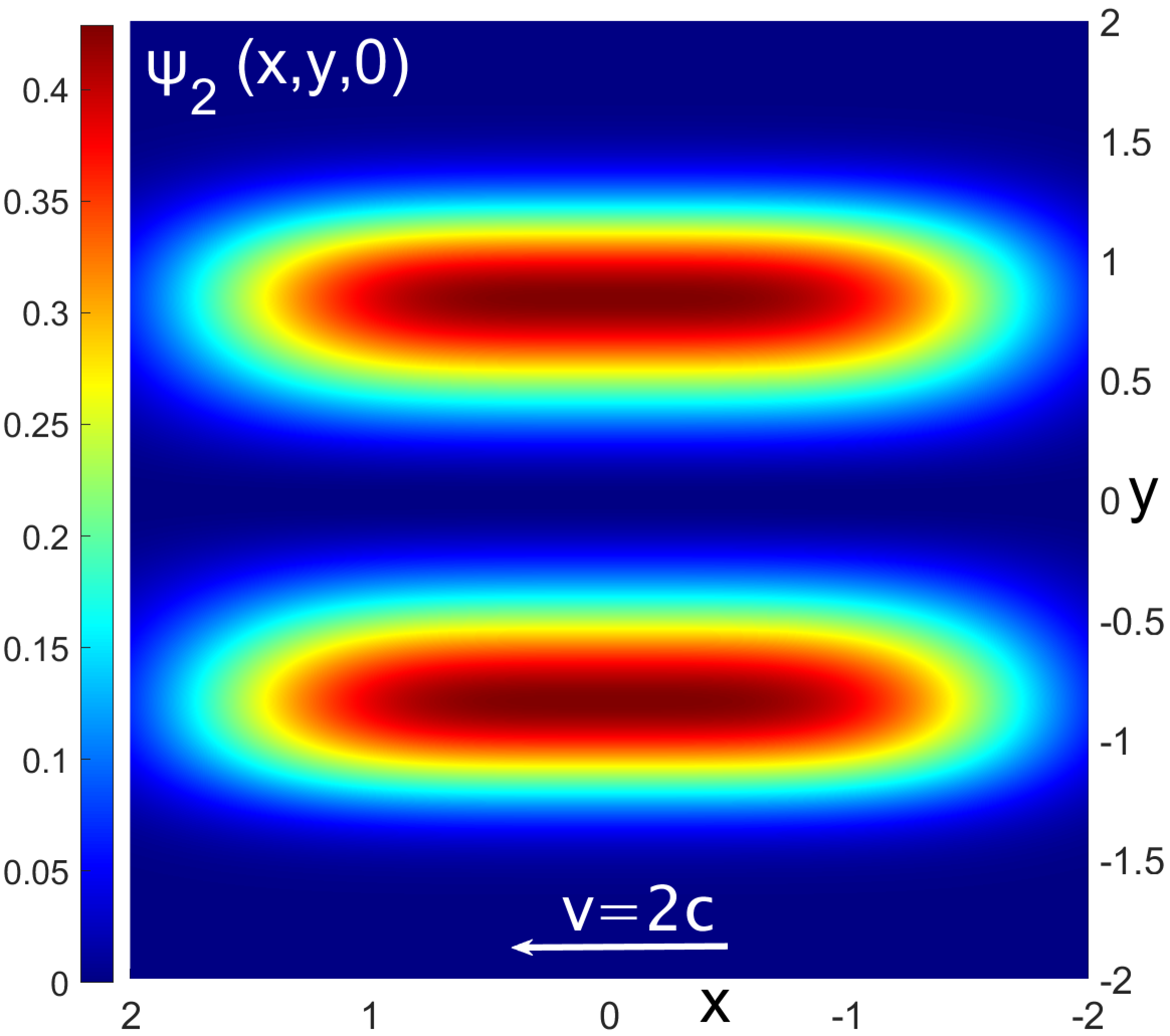}
		
	\end{tabular}
	\caption{Plots~a-f represent the functions (\ref{et})-(\ref{et2}) at $z=0$ at different speeds in the $x$-direction. }	\label{atrest}
\end{figure}

\begin{figure}[htp]
	
	\centering

	\begin{tabular}{cc}

		\includegraphics[width=65mm]{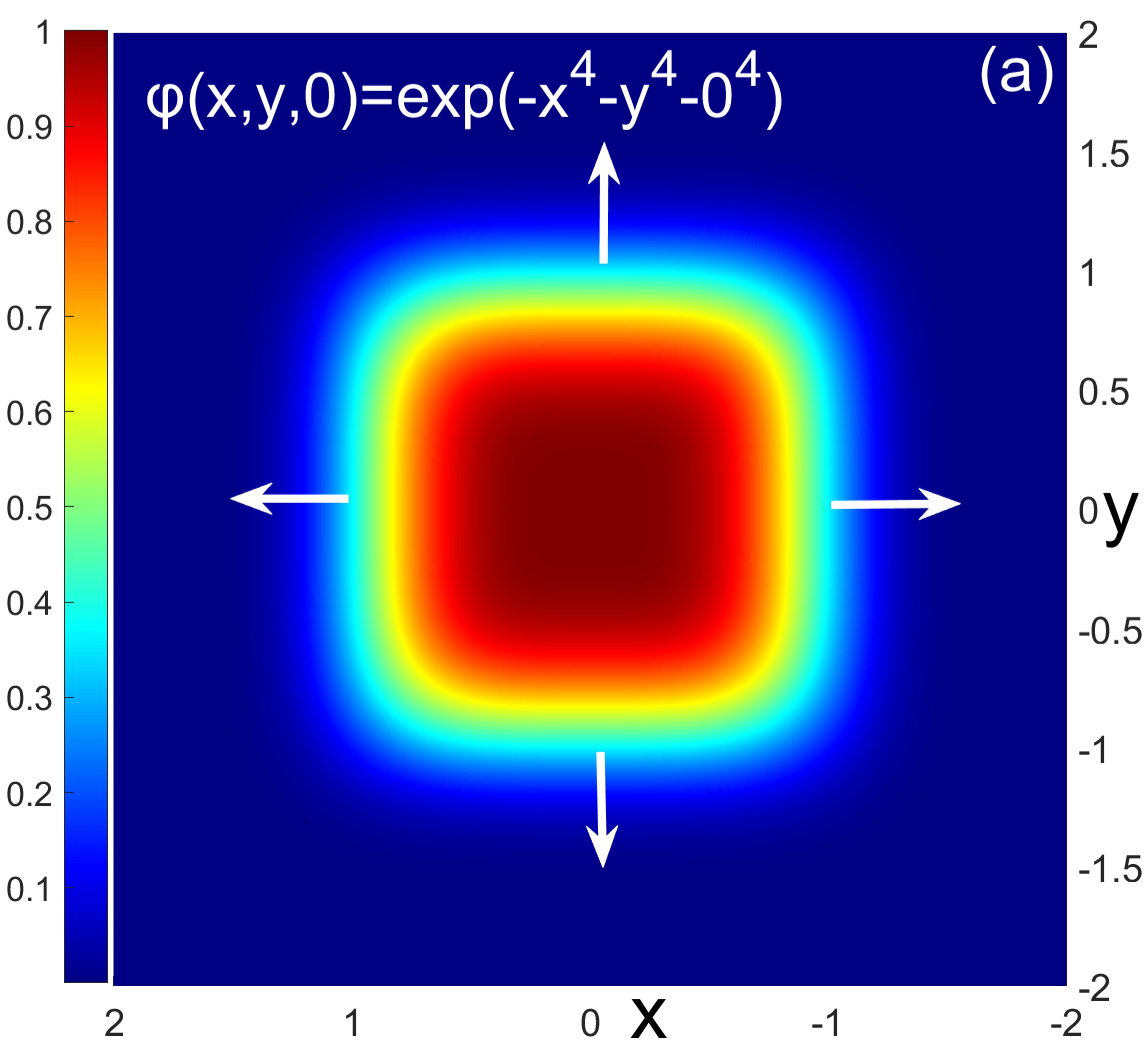}
		
		\includegraphics[width=65mm]{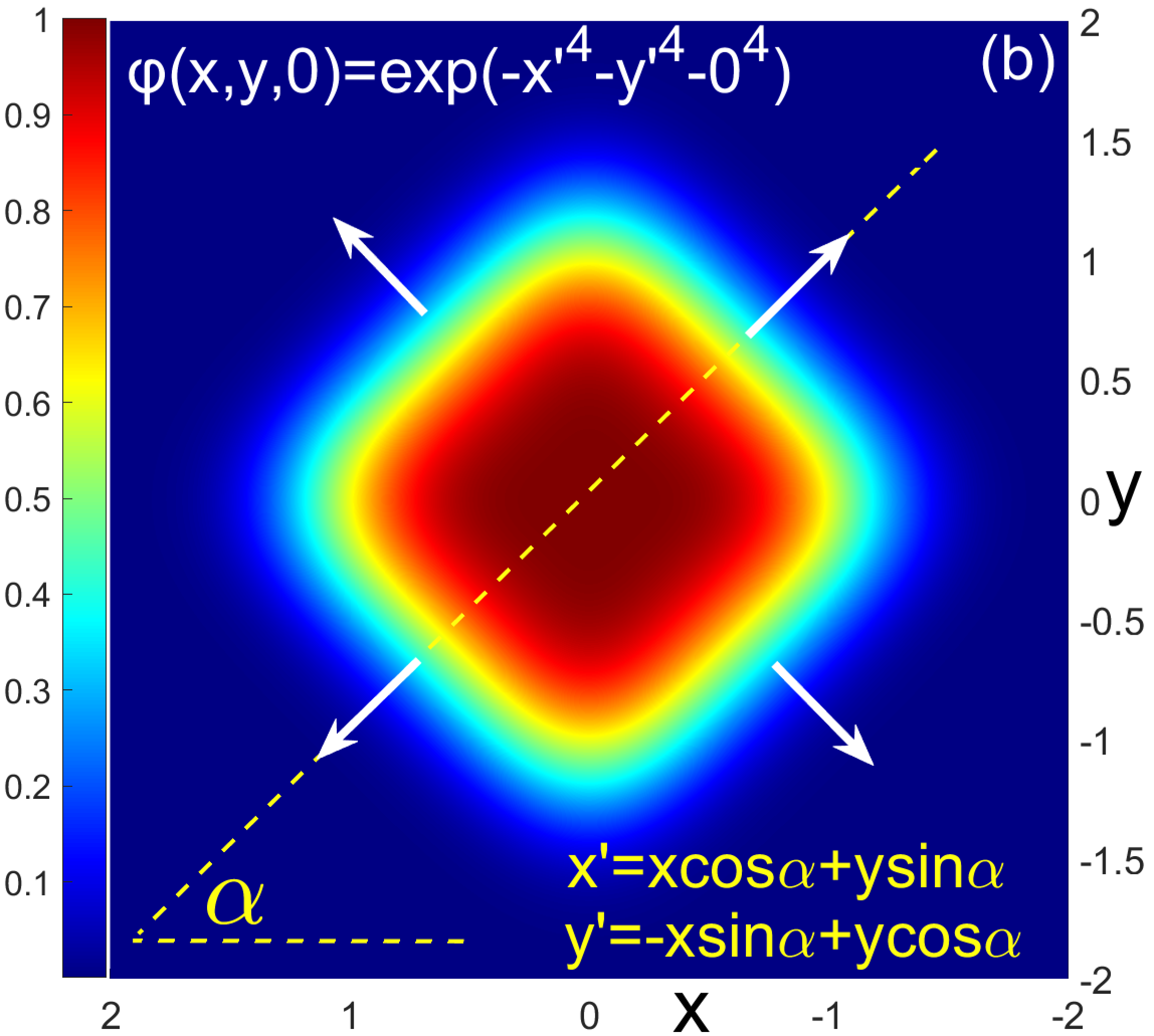}

	\end{tabular}
	\caption{The right directions (white arrows) in the ($x-y$)-plane for functions $\varphi(x,y,z)=e^{-x^4-y^4-z^4}$ and $\varphi(x,y,z)=e^{-x'^4-y'^4-z^4}$ where the FTL boosts can be performed. }\label{dirc}
\end{figure}

In the standard relativistic theory of classical fields, everything starts with introducing a suitable  Lagrangian density, and then its solutions are investigated. 
Contrary to the usual procedure, we can first consider a   solution of an unknown field model and then try to find a  Lagrangian density that supports it.   
For example, to have a zero-energy FTL solitary wave solution, we can propose four real scalar fields with  suitable formats at rest as follows:
\begin{eqnarray} \label{et}
	&& \Phi_{1}=\varphi=\exp(-x^4-y^4-z^4),\\ \label{et1}
	&&\Phi_{2}=\psi_{1}=x^2\exp(-x^4-y^4-z^4),\\ \label{et2} 
	&&\Phi_{3}=\psi_{2}=y^2\exp(-x^4-y^4-z^4),\\ \label{et3}  
	&&\Phi_{4}=\psi_{3}=z^2\exp(-x^4-y^4-z^4),
\end{eqnarray}
which can be considered as a solution of a $k$-field model in $3+1$ dimensions  with an unknown Lagrangian density\footnote{It is better to clarify in advance that  the set of  functions (\ref{et})-(\ref{et3}), (\ref{etx})-(\ref{et3x}), and (\ref{etr})-(\ref{et3r}), along with $\Theta=K_{\mu}x^{\mu}$, represent  different versions of a  special solution to the dynamical equations (\ref{jkt})-(\ref{bb2}). In other words, functionals (\ref{h1})-(\ref{h13}) and (\ref{jj})-(\ref{jj7})    all vanish simultaneously  for any set of these functions. 
}.  
Of course, we will show later that the stability considerations are required  to introduce the fifth field $\Phi_{o}=\Theta$ as well, which behaves like a quasi-free field and is not  important to consider here. The reason why we need  $5$ scalar fields will be clarified later. 
For FTL velocities in the $x$, $y$ and $z$-directions, functions (\ref{et})-(\ref{et3}) remain  localized real ones  as we expected. Namely, for motion in the  $x$-direction (see Fig.~\ref{atrest}), we have 
\begin{eqnarray} \label{etx}
	&& \varphi=\exp(-\Gamma^4(x-vt)^4-y^4-z^4),\\ \label{et1x}
	&&\psi_{1}=-\Gamma^2(x-vt)^2\exp(-\Gamma^4(x-vt)^4-y^4-z^4),\\ \label{et2x} 
	&&\psi_{2}=y^2\exp(-\Gamma^4(x-vt)^4-y^4-z^4),\\ \label{et3x}  
	&&\psi_{3}=z^2\exp(-\Gamma^4(x-vt)^4-y^4-z^4).
\end{eqnarray}
In other directions,   applying Lorentz transformations to FTL velocities does not lead to favorable results. These functions will either  no longer remain localized  or will not be real. However,  a  Poincar\'{e} invariant field model will be constructed in which any arbitrary  spatially  rotated version of functions (\ref{et})-(\ref{et})  can be an equivalent solution. For example, instead of Eqs. (\ref{et})-(\ref{et}),  we can perform  any  rotation about the $z$-axis:
\begin{eqnarray} \label{etr}
	&& \varphi=\exp(-x'^4-y'^4-z^4),\\ \label{et1r}
	&&\psi_{1}=x'^2\exp(-x'^4-y'^4-z^4),\\ \label{et2r} 
	&&\psi_{2}=y'^2\exp(-x'^4-y'^4-z^4),\\ \label{et3r}  
	&&\psi_{3}=z^2\exp(-x'^4-y'^4-z^4).
\end{eqnarray}
where $x'=(\cos(\alpha)x+\sin(\alpha)y)$ and $y'=(-\sin(\alpha)x+\cos(\alpha)y)$. Hence,  functions (\ref{etr})-(\ref{et3r}) are suitable  for FTL boosts along the $z$, $x'$, and $y'$-axis (see Fig.~\ref{dirc}). 

%Figure 2 for function 3 shows all possible directions for super-light velocities that end in a correct answer.

%Only for motion in the the $x$, $y$, and $z$ directions, these functions  have a suitable form for FTL velocities.   There are some considerations regarding these four introduced functions, but in the present situation, our primary goal is to introduce a $k$-field model for which the set of above functions is a zero-energy solitary wave solution. 

%We could not find a relativistic model with a standard Lagrangian density for which there is a solution like functions  (\ref{et})-(\ref{et3}). Nevertheless, we could find a relativistic $k$-field model for  which the set of functions (\ref{et})-(\ref{et3}) is a  zero-energy solution. To make such a model, we need such $5$ scalar fields $\varphi$, $\Theta$, and $\psi_{i}$ ($i=1,2,3$). The reason why we need  $5$ scalar fields will be explained later. 

For  scalar fields $\Theta$, $\varphi$ and $\psi_{i}$ ($i=1,2,3$), many independent  scalar functionals  can be introduced, including the following: 
 \begin{eqnarray} \label{h1}
	&& \mathbb{S}_{0}=\partial_{\mu}\Theta\partial^{\mu}\Theta-\Omega_{s}^2, 	
	\\&&\label{h0}
	\mathbb{S}_{1}=\partial_{\mu}\varphi\partial^{\mu}\varphi+16\varphi^{-1}(\psi_{1}^3+\psi_{2}^3+\psi_{3}^3),
	\\&&\label{h2}
	\mathbb{S}_{2}=\partial_{\mu}\psi_{1}\partial^{\mu}\psi_{1}+16\psi_{1}\varphi^{-3}[(1/4)\varphi^{4}+\psi_{1}^4-\varphi^2\psi_{1}^{2}+\psi_{1}(\psi_{2}^3+\psi_{3}^3)],
	\\&&\label{h3}
	\mathbb{S}_{3}=\partial_{\mu}\psi_{2}\partial^{\mu}\psi_{2}+16\psi_{2}\varphi^{-3}[(1/4)\varphi^{4}+\psi_{2}^4-\varphi^2\psi_{2}^{2}+\psi_{2}(\psi_{1}^3+\psi_{3}^3)],
	\\&&\label{h4}
	\mathbb{S}_{4}=\partial_{\mu}\psi_{3}\partial^{\mu}\psi_{3}+16\psi_{3}\varphi^{-3}[(1/4)\varphi^{4}+\psi_{3}^4-\varphi^2\psi_{3}^{2}+\psi_{3}(\psi_{1}^3+\psi_{2}^3)],
	\\&&\label{h5}
	\mathbb{S}_{5}=\partial_{\mu}\psi_{1}\partial^{\mu}\psi_{2}+16\psi_{1}\psi_{2}\varphi^{-3}[\psi_{1}^3+\psi_{2}^3+\psi_{3}^3-(1/2)\varphi^2(\psi_{1}+\psi_{2})],
	\\&& \mathbb{S}_{6}=\partial_{\mu}\psi_{1}\partial^{\mu}\psi_{3}+16\psi_{1}\psi_{3}\varphi^{-3}[\psi_{1}^3+\psi_{2}^3+\psi_{3}^3-(1/2)\varphi^2(\psi_{1}+\psi_{3})],
	\\&&\label{h7}
	\mathbb{S}_{7}=\partial_{\mu}\psi_{2}\partial^{\mu}\psi_{3}+16\psi_{2}\psi_{3}\varphi^{-3}[\psi_{1}^3+\psi_{2}^3+\psi_{3}^3-(1/2)\varphi^2(\psi_{2}+\psi_{3})],	
	\\&&\label{h8}
	\mathbb{S}_{8}~=\partial_{\mu}\varphi\partial^{\mu}\psi_{1}+16\psi_{1}\varphi^{-2}[\psi_{1}^3+\psi_{2}^3+\psi_{3}^3-(1/2)\psi_{1}\varphi^{2}],
	\\&&\label{h82}
	\mathbb{S}_{9}=\partial_{\mu}\varphi\partial^{\mu}\psi_{2}+16\psi_{2}\varphi^{-2}[\psi_{1}^3+\psi_{2}^3+\psi_{3}^3-(1/2)\psi_{2}\varphi^{2}],
	\\&&\label{h9}
	\mathbb{S}_{10}=\partial_{\mu}\varphi\partial^{\mu}\psi_{3}+16\psi_{3}\varphi^{-2}[\psi_{1}^3+\psi_{2}^3+\psi_{3}^3-(1/2)\psi_{3}\varphi^{2}],
	\\&&\label{h10}
	\mathbb{S}_{11}=\partial_{\mu}\varphi\partial^{\mu}\varphi~+16\varphi^{-1}[(\psi_{1}^3+\psi_{2}^3+\psi_{3}^3)+\varphi^3\ln(\varphi)+\varphi(\psi_{1}^2+\psi_{2}^2+\psi_{3}^2)],
	\\&&\label{h11}
	\mathbb{S}_{12}=\partial_{\mu}\varphi\partial^{\mu}\psi_{1}+16\psi_{1}\varphi^{-2}[\psi_{1}^3+\psi_{2}^3+\psi_{3}^3]+8[\psi_{2}^2+\psi_{3}^2+\varphi^2\ln(\varphi)],
	\\&&\label{h12}
	\mathbb{S}_{13}=\partial_{\mu}\varphi\partial^{\mu}\psi_{2}+16\psi_{2}\varphi^{-2}[\psi_{1}^3+\psi_{2}^3+\psi_{3}^3]+8[\psi_{1}^2+\psi_{3}^2+\varphi^2\ln(\varphi)],
	\\&&\label{h13}
	\mathbb{S}_{14}=\partial_{\mu}\varphi\partial^{\mu}\psi_{3}+16\psi_{3}\varphi^{-2}[\psi_{1}^3+\psi_{2}^3+\psi_{3}^3]+8[\psi_{1}^2+\psi_{2}^2+\varphi^2\ln(\varphi)].
\end{eqnarray}
These scalar functionals, except $\mathbb{S}_{0}$, are defined so that they are all zero simultaneously  for the set of functions (\ref{et})-(\ref{et3}) (or (\ref{etr})-(\ref{et3r})) and any relativistic boosted versions of them (e.g., functions (\ref{etx})-(\ref{et3x})).
According to formula (\ref{sf0}), all of the above scalars can be used to build a proper Lagrangian density with a  zero-energy solution. 
However, for this article,   11 of them (i.e., $\mathbb{S}_{1},\mathbb{S}_{1},\cdots,\mathbb{S}_{11}$) will be used. 
Each of the conditions $\mathbb{S}_{i}=0$ ($i=1,\cdots,11$) can be assumed as a PDE equation; thus, we have $11$ nonlinear coupled PDEs for four scalar fields $\varphi$ and $\psi_{i}$ ($i=1,2,3$). Hence, it is  fairly  normal to have no joint solution  other than the set of functions (\ref{et})-(\ref{et3}), which we initially wanted to be a joint solution. In fact,  the number of  independent PDEs $\mathbb{S}_{i}=0$ ($i>0$) must be larger than the number of fields to ensure that they have a unique joint solution.   Therefore, 5 of the PDEs  $\mathbb{S}_{i}=0$ ($i>0$)  seem to be sufficient, but to be more sure that there is only a unique  joint solution, 11 of them are arbitrarily chosen that are simpler. 
Of course, the scalar functional $\mathbb{S}_{11}$ must be selected because,  for the previous  conditions  $\mathbb{S}_{i}=0$ ($i=1,\cdots,10$), there is a continuous  range of solutions (scaling symmetry): 
\begin{eqnarray} \label{di}
	&& \varphi=(1+\xi)\exp(-x^4-y^4-z^4),\nonumber \\ 
	&& \psi_{i}=(1+\xi)(x^{i})^2\exp(-x^4-y^4-z^4),\quad (i=1,2,3),
\end{eqnarray}
where, $x^{1}=x,x^{2}=y$, $x^{3}=z$, and $\xi>-1$. To break this symmetry and have a unique joint solution, it is necessary to consider $\mathbb{S}_{11}$. If  one succeeds  in finding another joint solution for conditions $\mathbb{S}_{i}=0$ ($i=1,\cdots,11$), it would be
possible to select more new scalar functionals $\mathbb{S}_{i}$ ($i=12,13,\cdots$)  with additional restrictive conditions $\mathbb{S}_{i}=0$ to ensure the uniqueness of a joint solution.

To summarize, we are looking for a $k$-field model based on five scalar fields $\Theta$, $\varphi$ and $\Psi_{i}$ ($i=1,2,3$)  that has a single zero-energy  solution (at rest) in the following form
\begin{eqnarray} \label{et22}
	&& \Theta=K_{\mu}x^{\mu},\quad\quad\varphi=\exp(-x^4-y^4-z^4),\nonumber \\ 
	&& \psi_{i}=(x^{i})^2\exp(-x^4-y^4-z^4),\quad (i=1,2,3),
\end{eqnarray}
where $x^{0}=ct$. It should be noted that for $\Theta=K_{\mu}x^{\mu}$ we have  $\mathbb{S}_{0}=0$, provided  $K_{\mu}K^{\mu}=\Omega_{s}^2$.  A solitary wave solution is energetically stable if any arbitrary deformation above its background increases the total energy.  In relation to a zero-energy solution, energetically  stability condition imposes  severe restrictions on  series (\ref{sf0}). One way to ensure the energetical stability of a zero-energy solution is to look for special combinations in series (\ref{sf0}) that lead to energy density  functions in which all terms are positive definite.
In this regard, we can introduce a proper Lagrangian density as follow
 \begin{equation} \label{kk}
	{\cal L}=B\sum_{i=0}^{11}{\cal K}_{i}^3,
\end{equation}
where $B$ is a positive number, and
\begin{eqnarray} \label{jj}
	&&{\cal K}_{0}~=\varphi^2\mathbb{S}_{0},   \\&&\label{jj1}
	{\cal K}_{i}~=\varphi^2(b_{i}^2\mathbb{S}_{0}+\mathbb{S}_{i}), \quad~~(i=1,2,3,4,11), \\&&\label{jj2}
	{\cal K}_{5}~=\varphi^2(b_{5}^2\mathbb{S}_{0}+\mathbb{S}_{2}+\mathbb{S}_{3}+2\mathbb{S}_{5}), \\&&\label{jj3}
	{\cal K}_{6}~=\varphi^2(b_{6}^2\mathbb{S}_{0}+\mathbb{S}_{2}+\mathbb{S}_{4}+2\mathbb{S}_{6}), \\&&\label{jj4}
	{\cal K}_{7}~=\varphi^2(b_{7}^2\mathbb{S}_{0}+\mathbb{S}_{3}+\mathbb{S}_{4}+2\mathbb{S}_{7}), \\&&\label{jj5}
	{\cal K}_{8}~=\varphi^2(b_{8}^2\mathbb{S}_{0}+\mathbb{S}_{1}+\mathbb{S}_{2}+2\mathbb{S}_{8}), \\&&\label{jj6}
	{\cal K}_{9}~=\varphi^2(b_{9}^2\mathbb{S}_{0}+\mathbb{S}_{1}+\mathbb{S}_{3}+2\mathbb{S}_{9}), \\&&\label{jj7}
	{\cal K}_{10}=\varphi^2(b_{10}^2\mathbb{S}_{0}+\mathbb{S}_{1}+\mathbb{S}_{4}+2\mathbb{S}_{10}),
\end{eqnarray}
in which 
\begin{eqnarray} \label{ff}
	&&b_{i}=(1/\sqrt{2}\Omega_{s})(h_{i}+1), \quad (i=1,2,3,4,11),  \\&&\label{ff1} 
	b_{i}=(1/\sqrt{2}\Omega_{s})(\sigma_{i}+1), \quad (i=5,\cdots,10),  \\&&\label{ff11}	
	\sigma_{5}=h_{2}+h_{3}+2h_{5}, \\&&\label{ff3}
	\sigma_{6}=h_{2}+h_{4}+2h_{6}, \\&&\label{ff4}
	\sigma_{7}=h_{3}+h_{4}+2h_{7}, \\&&\label{ff5}
	\sigma_{8}=h_{1}+h_{2}+2h_{8}, \\&&\label{ff6}
	\sigma_{9}=h_{1}+h_{3}+2h_{9}, \\&&\label{ff7}
	\sigma_{10}=h_{1}+h_{4}+2h_{10},  
\end{eqnarray}
and
\begin{eqnarray} \label{hi}
	&&h_{1}=16\varphi^{-1}(\psi_{1}^3+\psi_{2}^3+\psi_{3}^3),  \\&&\label{hi2}
	h_{2}= 16\psi_{1}\varphi^{-3}[(1/4)\varphi^{4}+\psi_{1}^4-\varphi^2\psi_{1}^{2}+\psi_{1}(\psi_{2}^3+\psi_{3}^3)], \\&&\label{hi3}
	h_{3}=16\psi_{2}\varphi^{-3}[(1/4)\varphi^{4}+\psi_{2}^4-\varphi^2\psi_{2}^{2}+\psi_{2}(\psi_{1}^3+\psi_{3}^3)],\\&&\label{hi4}
	h_{4}=16\psi_{3}\varphi^{-3}[(1/4)\varphi^{4}+\psi_{3}^4-\varphi^2\psi_{3}^{2}+\psi_{3}(\psi_{1}^3+\psi_{2}^3)],\\&&\label{hi5}
	h_{5}=16\psi_{1}\psi_{2}\varphi^{-3}[\psi_{1}^3+\psi_{2}^3+\psi_{3}^3-(1/2)\varphi^2(\psi_{1}+\psi_{2})],\\&&\label{hi6}
	h_{6}=16\psi_{1}\psi_{3}\varphi^{-3}[\psi_{1}^3+\psi_{2}^3+\psi_{3}^3-(1/2)\varphi^2(\psi_{1}+\psi_{3})],\\&&\label{hi7}
	h_{7}=16\psi_{2}\psi_{3}\varphi^{-3}[\psi_{1}^3+\psi_{2}^3+\psi_{3}^3-(1/2)\varphi^2(\psi_{2}+\psi_{3})],\\&&\label{hi8}
	h_{8}=16\psi_{1}\varphi^{-2}[\psi_{1}^3+\psi_{2}^3+\psi_{3}^3-(1/2)\psi_{1}\varphi^{2}],
	\\&&\label{hi9}
	h_{9}=16\psi_{2}\varphi^{-2}[\psi_{1}^3+\psi_{2}^3+\psi_{3}^3-(1/2)\psi_{2}\varphi^{2}],
	\\&&\label{hi10}
	h_{10}=16\psi_{3}\varphi^{-2}[\psi_{1}^3+\psi_{2}^3+\psi_{3}^3-(1/2)\psi_{3}\varphi^{2}],
	\\&&\label{hi11}
	h_{11}=16\varphi^{-1}[(\psi_{1}^3+\psi_{2}^3+\psi_{3}^3)+\varphi^3\ln(\varphi)+\varphi(\psi_{1}^2+\psi_{2}^2+\psi_{3}^2)].
\end{eqnarray}
See  Eqs.~(\ref{h0})-(\ref{h10}) to understand how  we determine functions $h_{1},\cdots,h_{11}$. Since ${\cal K}_{i}$'s are linear combinations of $\mathbb{S}_{i}$'s, they  are all zero simultaneously for the special solution (\ref{et22}) as well. In fact, substituting $\mathbb{S}_{i}$'s  in the  Lagrangian density (\ref{kk}) leads to a form   according to formula (\ref{sf0}).
For system (\ref{kk}), one can obtain  the dynamical equations  easily:
\begin{eqnarray} \label{jkt}
	&&\partial_{\mu}\left(\sum_{i=0}^{11}{\cal K}_{i}^2  \left(\frac{\partial{\cal K}_{i}}{\partial(\partial_{\mu}\Theta)}\right)\right)=\sum_{i=0}^{11} {\cal K}_{i}\left[2(\partial_{\mu}{\cal K}_{i})   \frac{\partial{\cal K}_{i}}{\partial(\partial_{\mu}\Theta)}    +   {\cal K}_{i}\partial_{\mu}\left(\frac{\partial{\cal K}_{i}}{\partial(\partial_{\mu}\Theta)}\right)       \right]=0,\quad\quad\\&&\label{bb1}
	\sum_{i=0}^{11}  {\cal K}_{i}\left[2(\partial_{\mu}{\cal K}_{i})   \frac{\partial{\cal K}_{i}}{\partial(\partial_{\mu} \varphi)}    +   {\cal K}_{i}\partial_{\mu}\left(\frac{\partial{\cal K}_{i}}{\partial(\partial_{\mu} \varphi)}\right)    -    {\cal K}_{i}\frac{\partial{\cal K}_{i}}{\partial  \varphi}  \right] =0,\quad\quad\\&&\label{bb2}
	\sum_{i=1}^{11} {\cal K}_{i}\left[2(\partial_{\mu}{\cal K}_{i})   \frac{\partial{\cal K}_{i}}{\partial(\partial_{\mu}\psi_{j})}    +   {\cal K}_{i}\partial_{\mu}\left(\frac{\partial{\cal K}_{i}}{\partial(\partial_{\mu}\psi_{j})}\right)    -    {\cal K}_{i}\frac{\partial{\cal K}_{i}}{\partial \psi_{j}}   \right]=0. \quad (j=1,2,3).
\end{eqnarray}
Each term in the  above equations contains an expression placed in brackets $[\cdots]$ and multiplied by ${\cal K}_{i}$. Therefore, it is evident  that Eq.~(\ref{et22}) is a solution of system (\ref{kk}).

The energy-density  function (\ref{e5b2}) belongs to the  Lagrangian-density (\ref{kk}) and  can be simplified to:
\begin{eqnarray} \label{nnmn}
	&&\epsilon=\sum_{i=0}^{11}\varepsilon_{i}=B\sum_{i=0}^{11}{\cal K}_{i}^{2}\left[3C_{i}
	-{\cal K}_{i}\right],
\end{eqnarray}
which are divided  into eleven   distinct  parts, in which
\begin{equation}\label{cof}
	C_{i}=\dfrac{\partial{\cal K}_{i}}{\partial \dot{\Theta}}\dot{\Theta}+\dfrac{\partial{\cal K}_{i}}{\partial \dot{\varphi}}\dot{\varphi}+\sum_{j=1}^{3}\dfrac{\partial{\cal K}_{i}}{\partial \dot{\psi_{j}}}\dot{\psi_{j}}=
	\begin{cases}
		2c^{-2}\varphi^2\dot{\Theta}^{2} & \text{i=0}
		\\
		2c^{-2}\varphi^2(\dot{\varphi}^{2}+b_{1}^{2}\dot{\Theta}^2)  & \text{i=1},
		\\2c^{-2}\varphi^2(\dot{\psi_{1}}^{2}+b_{2}^{2}\dot{\Theta}^2)
		& \text{i=2},
		\\2c^{-2}\varphi^2(\dot{\psi_{2}}^{2}+b_{3}^{2}\dot{\Theta}^2)
		& \text{i=3},
		\\2c^{-2}\varphi^2(\dot{\psi_{3}}^{2}+b_{4}^{2}\dot{\Theta}^2)
		& \text{i=4},
		\\2c^{-2}\varphi^2((\dot{\psi_{1}}+\dot{\psi_{2}})^{2}+b_{5}^{2}\dot{\Theta}^2)
		& \text{i=5},
		\\2c^{-2}\varphi^2((\dot{\psi_{1}}+\dot{\psi_{3}})^{2}+b_{6}^{2}\dot{\Theta}^2)
		& \text{i=6},
		\\2c^{-2}\varphi^2((\dot{\psi_{2}}+\dot{\psi_{3}})^{2}+b_{7}^{2}\dot{\Theta}^2)
		& \text{i=7},
		\\2c^{-2}\varphi^2((\dot{\varphi}+\dot{\psi_{1}})^{2}+b_{8}^{2}\dot{\Theta}^2)
		& \text{i=8},
		\\2c^{-2}\varphi^2((\dot{\varphi}+\dot{\psi_{2}})^{2}+b_{9}^{2}\dot{\Theta}^2)
		& \text{i=9},
		\\2c^{-2}\varphi^2((\dot{\varphi}+\dot{\psi_{3}})^{2}+b_{10}^{2}\dot{\Theta}^2)
		& \text{i=10},
		\\2c^{-2}\varphi^2(\dot{\varphi}^{2}+b_{11}^{2}\dot{\Theta}^2)
		& \text{i=11}.
	\end{cases}
\end{equation}
It should be noted  that $C_{i}$'s are all positive definite expressions.    After a straightforward calculation, we obtain:
\begin{eqnarray} \label{eis1}
	&&\varepsilon_{0}=B\varphi^2{\cal K}_{0}^2[5c^{-2}\dot{\Theta}^2+(\boldsymbol\nabla\Theta)^2+\Omega_{s}^2],\\ \label{eis2}&&
	\varepsilon_{1}=B\varphi^2{\cal K}_{1}^2[5c^{-2}(b_{1}^{2}\dot{\Theta}^2+\dot{\varphi}^2)+b_{1}^{2}(\boldsymbol\nabla\Theta)^2+(\boldsymbol\nabla \varphi)^2+(1/2)(h_{1}^2+1)],\quad\\ \label{eis3}&&
		\varepsilon_{2}=B\varphi^2{\cal K}_{2}^2[5c^{-2}(b_{2}^{2}\dot{\Theta}^2+\dot{\psi_{1}}^2)+b_{2}^{2}(\boldsymbol\nabla\Theta)^2+(\boldsymbol\nabla \psi_{1})^2+(1/2)(h_{2}^2+1)],\\ \label{eis4}&&
\varepsilon_{3}=B\varphi^2{\cal K}_{3}^2[5c^{-2}(b_{3}^{2}\dot{\Theta}^2+\dot{\psi_{2}}^2)+b_{3}^{2}(\boldsymbol\nabla\Theta)^2+(\boldsymbol\nabla \psi_{2})^2+(1/2)(h_{3}^2+1)],\\
	\label{eis5}&&
	\varepsilon_{4}=B\varphi^2{\cal K}_{4}^2[5c^{-2}(b_{4}^{2}\dot{\Theta}^2+\dot{\psi_{3}}^2)+b_{4}^{2}(\boldsymbol\nabla\Theta)^2+(\boldsymbol\nabla \psi_{3})^2+(1/2)(h_{4}^2+1)],\\
	\label{eis6}&&
	\varepsilon_{5}=B\varphi^2{\cal K}_{5}^2[5c^{-2}(b_{5}^{2}\dot{\Theta}^2+(\dot{\psi_{1}}+\dot{\psi_{2}})^2)+b_{5}^{2}(\boldsymbol\nabla\Theta)^2+(\boldsymbol\nabla \psi_{1}+\boldsymbol\nabla \psi_{2})^2+(1/2)(\sigma_{5}^2+1)],\quad\quad\\ \label{eis7}&&
	\varepsilon_{6}=B\varphi^2{\cal K}_{6}^2[5c^{-2}(b_{6}^{2}\dot{\Theta}^2+(\dot{\psi_{1}}+\dot{\psi_{3}})^2)+b_{6}^{2}(\boldsymbol\nabla\Theta)^2+(\boldsymbol\nabla \psi_{1}+\boldsymbol\nabla \psi_{3})^2+(1/2)(\sigma_{6}^2+1)],\quad\quad \\ \label{eis8}&&
		\varepsilon_{7}=B\varphi^2{\cal K}_{7}^2[5c^{-2}(b_{7}^{2}\dot{\Theta}^2+(\dot{\psi_{2}}+\dot{\psi_{3}})^2)+b_{7}^{2}(\boldsymbol\nabla\Theta)^2+(\boldsymbol\nabla \psi_{2}+\boldsymbol\nabla \psi_{3})^2+(1/2)(\sigma_{7}^2+1)],\quad\quad \\ \label{eis9}&&
		\varepsilon_{8}=B\varphi^2{\cal K}_{8}^2[5c^{-2}(b_{8}^{2}\dot{\Theta}^2+(\dot{\varphi}+\dot{\psi_{1}})^2)+b_{8}^{2}(\boldsymbol\nabla\Theta)^2+(\boldsymbol\nabla \varphi+\boldsymbol\nabla \psi_{1})^2+(1/2)(\sigma_{8}^2+1)],\quad\quad \\ \label{eis10}
	&& \varepsilon_{9}=B\varphi^2{\cal K}_{9}^2[5c^{-2}(b_{9}^{2}\dot{\Theta}^2+(\dot{\varphi}+\dot{\psi_{2}})^2)+b_{9}^{2}(\boldsymbol\nabla\Theta)^2+(\boldsymbol\nabla \varphi+\boldsymbol\nabla \psi_{2})^2+(1/2)(\sigma_{9}^2+1)],\quad\quad \\ \label{eis11}
	&& \varepsilon_{10}=B\varphi^2{\cal K}_{10}^2[5c^{-2}(b_{10}^{2}\dot{\Theta}^2+(\dot{\varphi}+\dot{\psi_{3}})^2)+b_{10}^{2}(\boldsymbol\nabla\Theta)^2+(\boldsymbol\nabla \varphi+\boldsymbol\nabla \psi_{3})^2+(1/2)(\sigma_{10}^2+1)].\quad\quad
	\\ \label{eis14}
	&& \varepsilon_{11}=B\varphi^2{\cal K}_{11}^2[5c^{-2}(b_{11}^{2}\dot{\Theta}^2+\dot{\varphi}^2)+b_{11}^{2}(\boldsymbol\nabla\Theta)^2+(\boldsymbol\nabla \varphi)^2+(1/2)(h_{11}^2+1)],
\end{eqnarray}
Each of $\varepsilon_{i}$'s becomes zero for the special solution (\ref{et22}); thus,  it is a zero energy solution, as we expected.
Also, according to Eqs.~(\ref{eis1})-(\ref{eis14}),  all terms  in the energy density function are  positive definite; hence,  we are sure that  the single  zero-energy solution (\ref{et22})  is an energetically  stable entity.
More precisely, let us  consider the variation of the energy density function under any small deformations. Any arbitrary small  deformed version  of the zero-energy solution (\ref{et22}) can be introduced as follows:
\begin{eqnarray} \label{et22v}
	&& \Theta=K_{\mu}x^{\mu}+\delta \Theta,\quad\quad\varphi=\exp(-x^4-y^4-z^4)+\delta \varphi,\\ 
	&& \psi_{i}=(x^{i})^2\exp(-x^4-y^4-z^4)+\delta \psi_{i},\quad (i=1,2,3),\nonumber 
\end{eqnarray}
where $\delta \varphi$, $\delta \Theta$, and $\delta \psi_{j}$  (small variations) are  considered to be any arbitrary  small functions  of space-time. If one investigates  $\varepsilon_{i}$ ($i=1,\cdots,11$)  and keeps the terms to the least order of variations, it  yields:
\begin{eqnarray} \label{so4}
	&&\varepsilon_{i}+\delta\varepsilon_{i}=\delta\varepsilon_{i}=B[3(C_{i}+\delta C_{i})({\cal K}_{i}+\delta{\cal K}_{i})^{2}-({\cal K}_{i}+\delta{\cal K}_{i})^{3}] \approx [3BC_{i}(\delta{\cal K}_{i})^{2}]>0.\quad\quad
\end{eqnarray}
Thus, $\delta\epsilon=\sum_{i=0}^{11}\delta\varepsilon_{i}$, and then $\delta E=\int\delta\epsilon~ d^3x$ are always positive definite values for all small variations; that is, the zero-energy  solution (\ref{et22}) is energetically stable.
Note that, for the zero-energy  solution (\ref{et22}), ${\cal K}_{i}=0$ and $\varepsilon_{i}=0$ ($i=1,\cdots,11$). The above calculations are considered for a non-moving version of a zero-energy solution (\ref{et22}) without any changes; the same results can also be obtained for its moving version  (even for the canonical FTL boosted versions).

As noted above, having a non-moving solution leads to its moving version by applying a relativistic boost. For example, the FTL version of the special solution (\ref{et22}) in the $x$-direction would be
\begin{eqnarray} \label{ftls}
	&& \Theta=K_{\mu}x^{\mu},\quad\varphi=\exp(-\widetilde{x}^4-y^4-z^4),\quad \psi_{1}=(\widetilde{x})^2\exp(-\widetilde{x}^4-y^4-z^4)\nonumber \\ 
	&& \psi_{i}=(x^{i})^2\exp(-\widetilde{x}^4-y^4-z^4),\quad (i=2,3),
\end{eqnarray}
where $\widetilde{x}=i\Gamma(x-vt)$. The important point  here is that the boosting process is not done for the phase field $\Theta$. 
In fact, for this field, regardless  of the velocity, there are diverging solutions such as $\Theta=\Omega_{s}(ct)$, $\Theta=(1/\sqrt{3})\Omega_{s}(2ct-x)$,  $\Theta=(1/\sqrt{12})\Omega_{s}(4ct+2x)$, and so on, provided that they satisfy  the constraint $\mathbb{S}_{0}=\partial_{\mu}\Theta\partial^{\mu}\Theta-\Omega_{s}^2=0$. Hence, for the  zero-energy solution (\ref{et22}) or (\ref{ftls}), for which ${\cal K}_{i}=0$ or $\mathbb{S}_{i}=0$ ($i = 0, 1, \cdots,11$), the form of $\varphi$ and $\psi_{j}$ ($j=1,2,3$) are unique, but $\Theta$ can be considered as a quasi-free field.
In  the regions far enough  from a zero-energy particle-like solution (\ref{et22}), the scalar fields  $\varphi$, $\psi_{j}$ ($j=1,2,3$), and then   $\epsilon$ are almost zero;  thus, there is not any rigorous restriction on $\Theta$ to be in  the standard form  $\Theta=K_{\mu}x^{\mu}$   as  a solution of the condition $\mathbb{S}_{0}=0$ (i.e., it would be a completely free field). 
In fact, the phase field $\Theta$ (which can be called the catalyzer field) is only applied to have  a system for which all the terms in the energy density function are positive definite. In other words, it is applied to guarantee the stability of the single zero-energy solution (\ref{et22}).

\section{An FTL soliton solution with non-zero energy} \label{sec5}

This section shows how adding three new terms to the Lagrangian density (\ref{kk}) can lead to an FTL soliton solution with non-zero energy. We need to introduce a new phase field $\theta$, and, hence, the modified Lagrangian density is:
 \begin{equation} \label{kk2}
	{\cal L}=B\sum_{i=0}^{11}{\cal K}_{i}^3+B{\cal K}_{12}^3+B{\cal K}_{13}^3+\varphi^2\mathbb{S}_{13}=B\sum_{i=0}^{13}{\cal K}_{i}^3+\varphi^2\mathbb{S}_{13},
\end{equation}
where
\begin{eqnarray} \label{ss}
	&&{\cal K}_{12}~=\varphi^2(b_{12}^2\mathbb{S}_{0}+\mathbb{S}_{1}+\mathbb{S}_{13}+2\mathbb{S}_{12}),    \\&&\label{ss1}
	{\cal K}_{13}~=\varphi^2(\mathbb{S}_{0}+\mathbb{S}_{13}),
	 \\&&\label{ss2}
	\mathbb{S}_{12}=\partial_{\mu}\varphi\partial^{\mu}\theta,  \\&&\label{ss3}
	 \mathbb{S}_{13}=\partial_{\mu}\theta\partial^{\mu}\theta+\Omega_{s}^2, \\&&\label{ss4}
	b_{12}=(1/\sqrt{2}\Omega_{s})(h_{1}+\Omega_{s}^2+1).
\end{eqnarray}
The modified dynamical equations would be
\begin{eqnarray} \label{bbb}
	&&\partial_{\mu} J^{\mu}=\partial_{\mu}\left(\sum_{i=0}^{13}{\cal K}_{i}^2  \frac{\partial{\cal K}_{i}}{\partial(\partial_{\mu}\Theta)}\right)=0,\\&&\label{bbb1}
	3B\sum_{i=0}^{13}  {\cal K}_{i}\left[2(\partial_{\mu}{\cal K}_{i})   \frac{\partial{\cal K}_{i}}{\partial(\partial_{\mu} \varphi)}    +   {\cal K}_{i}\partial_{\mu}\left(\frac{\partial{\cal K}_{i}}{\partial(\partial_{\mu} \varphi)}\right)    -    {\cal K}_{i}\frac{\partial{\cal K}_{i}}{\partial  \varphi}  \right]-2\varphi \mathbb{S}_{13} =0,\quad\quad\\&&\label{bbb2}
	\sum_{i=1}^{11} {\cal K}_{i}\left[2(\partial_{\mu}{\cal K}_{i})   \frac{\partial{\cal K}_{i}}{\partial(\partial_{\mu}\psi_{j})}    +   {\cal K}_{i}\partial_{\mu}\left(\frac{\partial{\cal K}_{i}}{\partial(\partial_{\mu}\psi_{j})}\right)    -    {\cal K}_{i}\frac{\partial{\cal K}_{i}}{\partial \psi_{j}}   \right]=0. \quad (j=1,2,3).\quad \quad \\&&\label{bbb3}
	\partial_{\mu}j^{\mu}=\partial_{\mu}\left(\sum_{i=12}^{13} 3 B{\cal K}_{i}^2  \left(\frac{\partial{\cal K}_{i}}{\partial(\partial_{\mu}\theta)}\right)+2\varphi^2\partial^{\mu}\theta\right)=0,
\end{eqnarray} 
According to Eqs.~(\ref{bbb}) and (\ref{bbb3}), $J^{0}$ and $j^{0}$ are two conserved quantities. Regarding the energy density, we have to add three new terms to Eq.~(\ref{nnmn}), i.e.,
\begin{eqnarray} \label{nnmn2}
	&&\epsilon=\sum_{i=0}^{14}\varepsilon_{i}=B\sum_{i=0}^{13}{\cal K}_{i}^{2}\left[3C_{i}
	-{\cal K}_{i}\right]+\varepsilon_{14},
\end{eqnarray}
where
\begin{eqnarray} \label{eis111}
	&&\varepsilon_{12}=B\varphi^2{\cal K}_{12}^2[5c^{-2}(b_{12}^2\dot{\Theta}^2+(\dot{\varphi}+\dot{\theta})^2)+b_{12}^2(\boldsymbol\nabla\Theta)^2+(\boldsymbol\nabla\theta+\boldsymbol\nabla\varphi)^2+(1/2)((h_{1}+\Omega_{s}^2)^2+1)],\quad\quad\quad \\ \label{eis12}
	&&\varepsilon_{13}= B\varphi^2{\cal K}_{13}^2[5c^{-2}(\dot{\Theta}^2+\dot{\theta}^2)+(\boldsymbol\nabla\Theta)^2+(\boldsymbol\nabla\theta)^2].\quad\quad\\ \label{eis13}
	&&\varepsilon_{14}= \varphi^2[c^{-2}\dot{\theta}^2+(\boldsymbol\nabla\theta)^2-\Omega_{s}^2]
\end{eqnarray}

Here, similar to the previous section, we expect there to be a solution for which ${\cal K}_{i}$ or $\mathbb{S}_{i}$ ($i=1,\cdots13$) becomes zero. Despite such a solution,  Eqs.~(\ref{bbb})-(\ref{bbb2}) are all satisfied  automatically. Also, the first part of Eq.~(\ref{bbb3}) becomes zero automatically. Regarding the second part of equation Eq.~(\ref{bbb3}), we expect $\theta$ to be such that this part becomes zero, i.e., we expect 
 \begin{eqnarray} \label{pk}
\partial_{\mu}(\varphi^2\partial^{\mu}\theta)=\varphi(2\partial_{\mu}\varphi\partial^{\mu}\theta+\varphi\partial_{\mu}\partial^{\mu}\theta)=0
 \end{eqnarray}
Supposing an FTL solution moving along the $x$-axis, the space-time dependence  of the scalar field $\varphi$ is introduced according to Eq.~($\ref{etx}$). Therefore, if $\theta$ is expressed as  $\theta=k_{\mu}x^{\mu}=\omega t-kx$ ($k^{\mu}=(\omega/c,k,0,0)$),  to satisfy Eq.~(\ref{pk}),  condition $k=v\omega/c^2$ must be met between the wave number $k$ and the frequency $\omega$. Also, according to condition $\mathbb{S}_{13}=0$, we expect that $\omega$ and $k$ must satisfy the relation $k_{\mu}k^{\mu}=\omega^2/c^2-k^2=-\Omega_{s}^2$. Thus, by combining these two conditions, frequency $\omega$ is obtained:
 \begin{eqnarray} \label{vb}
\omega=\dfrac{\Omega_{s}c}{\sqrt{v^2/c^2-1}}=\Omega_{s}\Gamma c.
\end{eqnarray}

To summarize,  there is an FTL  solitary wave solution for dynamical equations (\ref{bbb})-(\ref{bbb3}) moving, for example, along  the $x$-axis   given as: 
\begin{eqnarray} \label{xx}
	&& \Theta=K_{\mu}x^{\mu},\quad\quad\theta=k_{\mu}x^{\mu}=\Omega_{s}\Gamma c(t- vx/c^2) ,\nonumber\\ 
	&& \varphi=\exp(-\Gamma^4(x-vt)^4-y^4-z^4),\nonumber\\ 
	&&\psi_{1}=-\Gamma^2(x-vt)^2\exp(-\Gamma^4(x-vt)^4-y^4-z^4),\nonumber\\ 
	&&\psi_{2}=y^2\exp(-\Gamma^4(x-vt)^4-y^4-z^4),\nonumber\\  
	&&\psi_{3}=z^2\exp(-\Gamma^4(x-vt)^4-y^4-z^4).
\end{eqnarray}
where   $K_{\mu}K^{\mu}=\Omega_{s}^2$.
For such an FTL solution  (\ref{xx}),  terms $\varepsilon_{1}$ to $\varepsilon_{13}$  become zero, and only $\varepsilon_{14}$ is non-zero. Thus, the energy of the particle-like solution (\ref{xx}) is obtained by spatial integration of $\varepsilon_{14}$ over the whole space:
\begin{eqnarray} \label{cc}
E=\int \varepsilon_{14} d^3x=2c^{-2}\omega^2\int \varphi^2 d^3x=2\Gamma\Omega_{s}^2\sigma=\dfrac{2\Omega_{s}^2\sigma}{\sqrt{v^2/c^2-1}}
\end{eqnarray}
where 
\begin{eqnarray} \label{xc}
	\sigma=\int_{-\infty}^{\infty}\int_{-\infty}^{\infty}\int_{-\infty}^{\infty} \exp{[-2(x^4+y^4+z^4)]} dxdydz=\dfrac{2^{\frac{3}{4}}\pi^3}{8[\Gamma'(3/4)]^3},
\end{eqnarray}
in which $\Gamma'$ is the Gamma function. Similarly, using Eq.~(\ref{e5a}), the momentum can be obtained  by 
\begin{eqnarray} \label{mo}
	\textbf{P}=P^{1}\hat{i}=\hat{i}c^{-1} \int T^{01} d^3x=-\hat{i} \int \Pi_{\theta}\frac{\partial \theta}{\partial x} d^3x=\hat{i} 2c^{-2}k\omega\int \varphi^2 d^3x=\frac{Ev}{c^2} \hat{i}.
\end{eqnarray}
where $\Pi_{\theta}=\frac{\partial{\cal L}}{\partial \dot{\theta}}$. The results obtained here are precisely according to the kinematic relations (\ref{stf}). Note that the other conjugate fields,  $\Pi_{\varphi}$, $\Pi_{\Theta}$,  $\Pi_{\Psi_{j}}$ ($j=1,2,3$),   are all zero for  the FTL solution  (\ref{xx}). 
Furthermore, it should be noted that using  the new field $\theta$ in the modified model (\ref{kk2}) makes solution (\ref{xx})  exist only for the FTL speeds, but its sublight version does not exist.
In fact,  conditions $k=v\omega/c^2$ and $k_{\mu}k^{\mu}=-\Omega_{s}^2$ cannot be satisfied simultaneously for sublight speeds ($v<c$).

If one assumes solution (\ref{xx}) as a  particle-like entity, its multi-particle-like version can be constructed due to the non-topological property of fields $\varphi$ and $\psi$. For example, if one considers solution (\ref{xx}) in $N$ different versions, each of which has its own initial location $x_{i}$ and  velocity $v_{i}$ along the $x$-axis,  the following summations would be approximately a multi-particle solution  at the initial times (the times that are close to $t = 0$): 
\begin{eqnarray} \label{mxx}
	&& \Phi=\sum_{i=1}^{N}\left[\exp(-\Gamma_{i}^4(x-v_{i}t-x_{i})^4-y^4-z^4)\right],\nonumber\\ 
	&&\Psi_{1}=-\sum_{i=1}^{N}\left[\Gamma_{i}^2(x-v_{i}t-x_{i})^2\exp(-\Gamma^4(x-v_{i}t-x_{i})^4-y^4-z^4)\right],\nonumber\\ 
	&&\Psi_{2}=\sum_{i=1}^{N}\left[y^2\exp(-\Gamma_{i}^4(x-v_{i}t-x_{i})^4-y^4-z^4)\right],\nonumber\\  
	&&\Psi_{3}=\sum_{i=1}^{N}\left[z^2\exp(-\Gamma_{i}^4(x-v_{i}t-x_{i})^4-y^4-z^4)\right],
\end{eqnarray}
provided $|x_{i+1}-x_{i}|$ are large enough so that the tail of each would be (almost) zero at the position of the others. The greater the distances between the non-topological solutions,
the more accurate this approximation will be.

The phase field $\theta$ in a multi-particle solution must change from one solution to another.
In regions between two separate FTL  solutions, the scalar fields $\varphi$ and $\psi_{j}$ ($j=1,2,3$) are zero everywhere; hence, there is no  restriction for $\theta$ to be a  solution of the conditions $\mathbb{S}_{12}=0$ and $\mathbb{S}_{13}=0$. To put it differently,  where the scalar fields  $\varphi$ and $\psi_{j}$ ($j=1,2,3$) are almost zero, the phase field $\theta$ is free and can evolve without any serious restriction. 
For example, for a two-particle-like solution with $v_{1}=2c$ and $v_{1}=3c$ along the $x$-axis, the phase field must change from $\theta=\frac{\Omega_{s}}{\sqrt{3}}c(t-2x/c)$ at the position of the first FTL solution to $\theta=\frac{\Omega_{s}}{\sqrt{8}}c(t-3x/c)$ at the position of the second one. In other words,  where the   scalar fields  $\varphi$ and $\psi_{j}$ ($j=1,2,3$) are almost zero, the phase field $\theta$ is completely free and evolves without any  restriction. There is a similar argument  for the phase field $\Theta$, but it is independent of  the speed of a single FTL  solution (\ref{xx}).  Thus, it may not change at all and  may have a definite form, e.g., $\Theta=\Omega_{s}ct$ as a solution of condition $\mathbb{S}_{0}=0$, throughout the space, independent of any number of particle-like solutions.

\section{Stability Considerations} \label{sec6}

We designed the model so that $\varepsilon_{i}$'s ($i=1,\cdots,13$) are positive definite functionals and all are zero for the single soliton solution (\ref{xx}). This means, as we showed in Eq.~(\ref{so4}), that any deformation above the background of soliton solution (\ref{xx}) leads to an increase in the total energy. However,  $\varepsilon_{14}$ leads to a different situation, i.e,  $\delta\varepsilon_{14}$ is not necessarily positive for all deformations:
\begin{eqnarray} \label{e14}
\delta\varepsilon_{14}= 2\varphi\delta\varphi[c^{-2}\dot{\theta}^2+(\boldsymbol\nabla\theta)^2-\Omega_{s}^2]+2\varphi^2[c^{-2}\dot{\theta}\delta\dot{\theta}+\delta(\boldsymbol\nabla\theta)\cdot \boldsymbol\nabla\theta].
\end{eqnarray}   
According  to Eq.~(\ref{so4}),  $\delta\varepsilon_{i}$'s ($i=1,\cdots,13$)  are  functionals of the second order of variations and are  positive definite expressions. However, $\delta\varepsilon_{14}$ is not positive definite and is only a functional  of the first order of variations $\delta\varphi$ and $\delta\theta$. 
Thus, when  $\delta\varphi=0$, $\delta\theta=0$, and for any non-zero variations $\delta\Theta$, $\delta\psi_{j}$ ($j=1,2,3$), the  energy density variation  $\delta\varepsilon$ and then the total energy variation $\delta E=\int \delta\epsilon$ always would be positive expressions. 
For small non-zero variations  $\delta\varphi$ and $\delta\theta$, we would typically  expect $|\delta\varphi|>|\delta\varphi|^2$ and $|\delta\theta|>|\delta\theta|^2$; thus, we  expect $|\delta\varepsilon_{14}|>\sum_{i=1}^{13}\delta\varepsilon_{i}$ as well. 
In other words,  the total energy variation  $\delta E$ may be  non-positive for such small non-zero deformations, i.e., the FTL solution (\ref{xx}) may not be energetically stable.
However, it should be
noted that all $\delta\varepsilon_{i}$ ($i=1,\cdots,13$) have parameter $B$, but $\delta\varepsilon_{14}$ does not; thus, a comparison between them needs
a more detailed investigation.

In general, the larger the value of parameter $B$, the inequality violation $|\delta\varepsilon_{14}|>\sum_{i=1}^{13}\delta\varepsilon_{i}$ occurs for smaller variations $\delta\varphi$ and $\delta\theta$.
 For example, for the case $B = 10^{40}$, the inequality $|\delta\varphi|>B|\delta\varphi|^2$ is
fulfilled only for the  small variations less than $10^{-20}$, which are not physically significant. Likewise, a similar comparison can be used between $|\delta\varepsilon_{14}|$ and  $\sum_{i=1}^{13}\delta\varepsilon_{i}$. This means the order of variations $\delta\varphi$ and $\delta\theta$ required for inequality $|\delta\varepsilon_{14}|>\sum_{i=1}^{13}\delta\varepsilon_{i}$ to be valid has an upper bound. In general,  parameter $B$ can be chosen so large that the violation of the energetical  stability condition (i.e., $\delta E>0$) of solution (\ref{xx}) occurs only for very small variations $\delta\varphi$ and  $\delta\theta$,  which can be ignored physically.

\begin{figure}[htp]
	
	\centering

	\begin{tabular}{cc}

		\includegraphics[width=48mm]{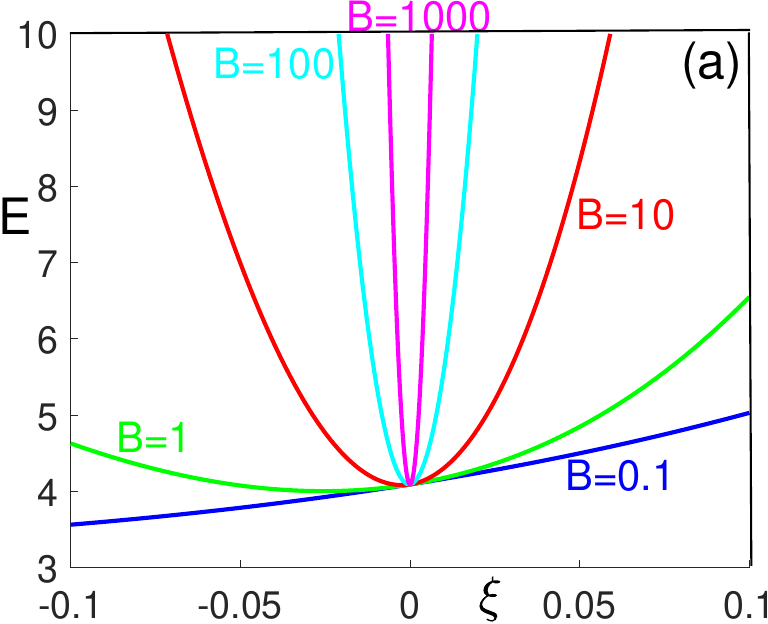}
		
		\includegraphics[width=48mm]{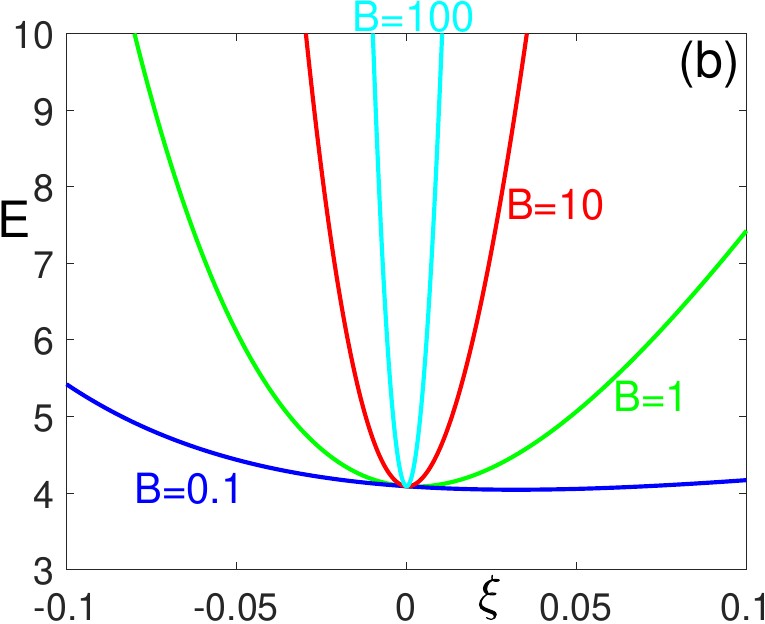}
		
		\includegraphics[width=48mm]{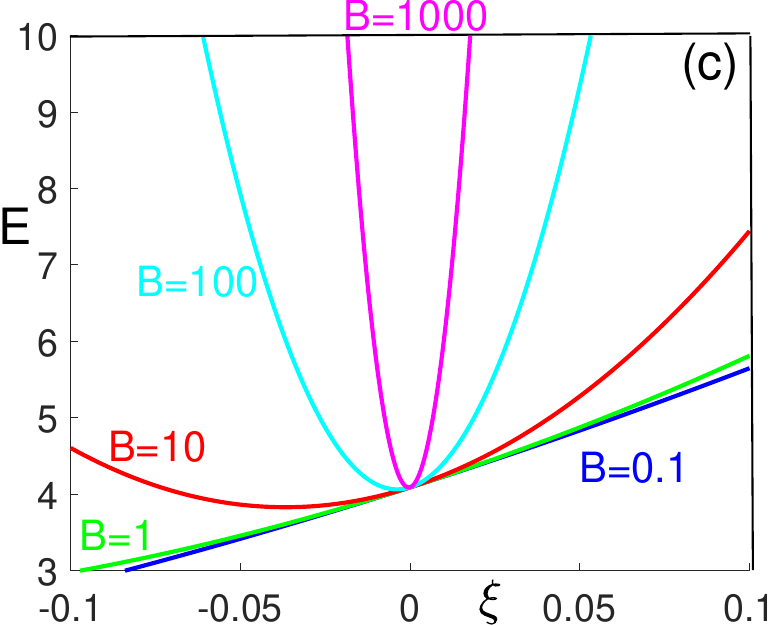}\\
		
		\includegraphics[width=48mm]{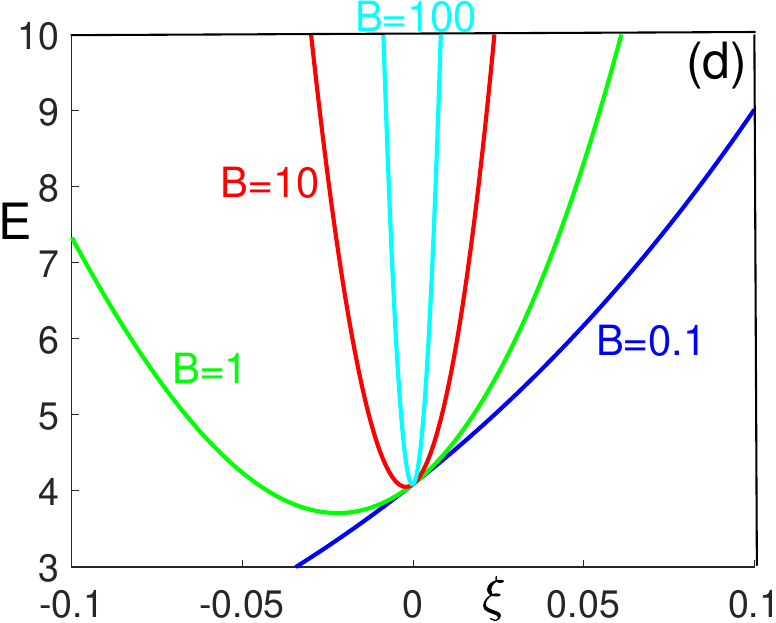}
		
		\includegraphics[width=48mm]{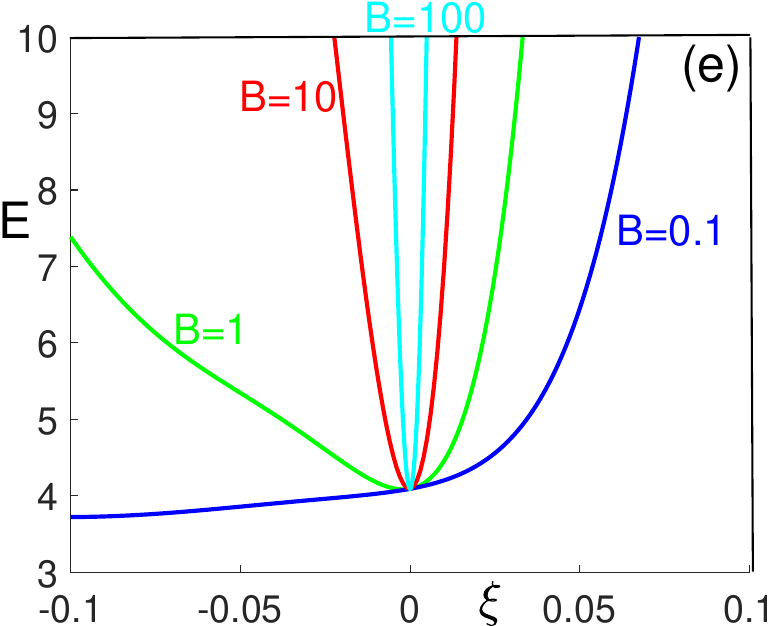}
		
		\includegraphics[width=48mm]{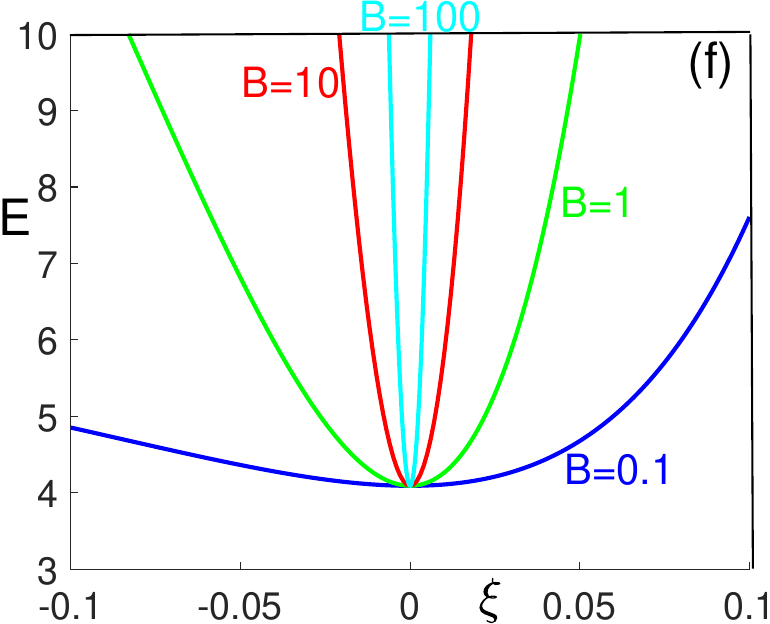}

	\end{tabular}
	\caption{Plots~a-f represent  variations of the total energy $E$ versus small $\xi$ for different deformations (\ref{adhoc})-(\ref{adhoc6})  at $t = 0$, respectively. Various
		color curves of  blue, green, red, cyan, and purple are related to $B=0.1$, $B=1$, $B=10$, $B=100$, and $B=1000$, respectively.}\label{var}
\end{figure}

To better see what was explained, by providing six arbitrary (ad hoc) small deformations, the variation of  total energy  can be considered numerically  for different values of  parameter $B$ (see Fig.~\ref{var}). For simplicity's sake, we set $c=1$, $v=2$, and $\Omega_{s}=1$; thus, six small ad hoc deformations can be introduced as follows: 
\begin{eqnarray} \label{adhoc}
	&&\theta+ \delta\theta=(\Gamma+\xi)t-(\Gamma v) x=k_{\mu}x^{\mu}+\xi t, \\ \label{adhoc2}
	&&\varphi+\delta\varphi=\exp((1+\xi)\zeta)=\exp(\zeta)+\sum_{n=0}^{\infty}\dfrac{(\xi\zeta)^n}{n!},\\ \label{adhoc3}
	&&\theta+ \delta\theta=(\Gamma+\xi)t-\sqrt{1+(\Gamma+\xi)^2}x,\\ \label{adhoc4}
	&&\theta+ \delta\theta=(\Gamma+\xi)t-((\Gamma+\xi)v)x=k_{\mu}x^{\mu}+\xi( t- v x),\\ \label{adhoc5}
	&&\begin{cases}
		\varphi+\delta\varphi=\exp{(\zeta)}(1+\xi),
		\\
		\psi_{1}+\delta\psi_{1}=-\Gamma^2(x-vt)^2 \exp{(\zeta)}(1+\xi),  
		\\\psi_{2}+\delta\psi_{2}=y^2 \exp{(\zeta)}(1+\xi),
		\\\psi_{3}+\delta\psi_{3}=z^2 \exp{(\zeta)}(1+\xi) ,
	\end{cases}
\\ \label{adhoc6}
&&\begin{cases}
	\psi_{2}+\delta\psi_{2}=y^2 \exp(\zeta)(1+\xi),
	\\\psi_{3}+\delta\psi_{3}=z^2 \exp(\zeta)(1+\xi),
\end{cases}
\end{eqnarray}
where $\zeta=(-\Gamma^4(x-vt)^4-y^4-z^4)$, and  $\xi$ is a small parameter, which can be considered as an indication of the order of small deformations. Note that the case $\xi=0$ leads to the same undeformed FTL soliton solution (\ref{xx}).
According to Fig.~(\ref{var}), it is clear that parameter $B$ plays an important role in increasing the stability of soliton solution (\ref{xx}). 
In other words,  the larger the value of $B$, the greater the increase in the total energy for any arbitrary small deformation above the background of the FTL  solution (\ref{xx}). 
Two new terms,  $B{\cal K}_{12}^3$ and $ B{\cal K}_{13}^3$, in the modified  Lagrangian density (\ref{kk2}) lead to the two constraints $\mathbb{S}_{12}=0$ and $\mathbb{S}_{13}=0$ on the field $\theta$. In fact, we considered such constraints to guarantee the least energy only for the solution  $\theta=k_{\mu}x^{\mu}=\Omega_{s}\Gamma c(t-xv/c^2)$ (provided that parameter $B$ is a large number).

\section{Summary and Conclusions} \label{sec7}

The hypothetical  FTL particles (tachyons) directly contradict the principle of causality and have never been observed experimentally. The existence of such particles may be rationalized if they have zero energy and momentum so that they have no observable interactions. Of course,  they may change the energy levels of a system, subject to conservation laws.

In this article, we showed that in the context of relativistic classical field theory,  FTL particle-like soliton solutions with zero energy are possible in $3+1$ dimensions;  that is, solutions for which the energy density vanishes  everywhere. First, we introduced the general form of Lagrangian densities (\ref{sf0}) that leads to  zero-energy solutions. 
Such Lagrangian densities should be expressed as powers of some scalar functionals, $\mathbb{S}_{i}$ ($i=1,2,3,\cdots$), all of which should become zero for a special solution (i.e., the same zero-energy solution). Second, the general rule for achieving an FTL solution based on a non-moving solution was explained as necessary. In particular, a model with a single zero-energy FTL soliton solution (\ref{ftls})  was introduced and fully investigated. This model was designed so that an infinite number scalar functionals (e.g., (\ref{h1})-(\ref{h13})) could be found for a proposed non-moving solution (\ref{et22}) that has turned into a  proper FTL solitary wave  solution (\ref{ftls}). Also, all terms in the corresponding energy density should be positive definite to guarantee the stability of the proposed FTL solitary wave  solution  (\ref{ftls}).

 Contrary to the principle of causality in reality, we showed that the classical relativistic field theory could even yield particle-like energetic FTL solutions in $3+1$ dimensions by presenting a modified model (\ref{kk2}). 
This modified model can be divided into two parts, one of which ($B\sum_{i=0}^{13}{\cal K}_{i}^3$) includes  parameter $B$ and results in zero energy for the FTL soliton solution (\ref{xx}), while  the other ($\varphi^2\mathbb{S}_{13}$) does not. 
We considered the stability of this energetic FTL solution against small perturbations. From a physical point of view, if the parameter $B$ is large,  stability is guaranteed. 
Also, this FTL solution (\ref{xx}) satisfies the same standard kinematic energy-momentum relations (\ref{stf}) obtained for hypothetical FTL particles in the general case, as  expected.

We believe the current study can help to complete our  vision for superluminal velocities. Indeed,  without using a new theory and only  based  on the same conventional relativistic classical field theory, we could know that  FTL particle-like solutions can exist. Why a fully relativistic model has such solutions and what the implications are can be an exciting topic for future investigations. Also, it may be possible to use models similar to the ones introduced in this article in gravity and cosmology.

\section*{Acknowledgement}

The author  wishes to express his appreciation to the Persian Gulf University Research Council for their constant support.

\end{document}